\documentclass{jfm}

\usepackage{graphicx}
\usepackage{natbib}
\usepackage{color}
\usepackage{amsmath}
\usepackage[usenames,dvipsnames]{xcolor}

\graphicspath{{FIGURES//}}

\title[Bluff body drag manipulation using pulsed jets and Coanda effect]{Bluff body drag manipulation \\using pulsed jets and Coanda effect}

\author[D. Barros, J. Bor\'{e}e, B. R. Noack, A. Spohn and T. Ruiz]%
{Diogo Barros$^{1,2}$%
  \thanks{Email address for correspondence: diogo.barros@ensma.fr},\ns
Jacques Bor\'{e}e$^1$,\ns
Bernd R. Noack$^{3,4}$\break
Andreas Spohn$^1$ and Tony Ruiz$^2$}

\affiliation{$^1$Institut Pprime, UPR-3346 CNRS -- Universit\'{e} de Poitiers -- ENSMA\\
Futuroscope Chasseneuil, 86360, France.\\[\affilskip]
$^2$PSA Peugeot-Citro\"{e}n, Centre Technique de V\'{e}lizy, V\'{e}lizy-Villacoublay, 78943, France.\\[\affilskip]
$^3$LIMSI -- CNRS, UPR 3251, Campus Universitare d'Orsay\\ Rue John Von Neumann, B\^at 508, F-91405 Orsay CEDEX, France.\\[\affilskip]
$^4$Institut f\"{u}r Str\"{o}mungsmechanik, Technische Universit\"{a}t Braunschweig\\ Hermann-Blenck-Stra{\ss}e 37, D-38108 Braunschweig, Germany.}

\pubyear{2010}
\volume{650}
\pagerange{119--126}
\date{?; revised ?; accepted ?. - To be entered by editorial office}
\begin{document}

\maketitle

\begin{abstract}

The impact of fluidic actuation on the wake and drag of a three-dimensional blunt body is investigated experimentally. 
Jets blowing tangentially to the main flow allow to force the wake with variable frequency and amplitude. 
Depending on the forcing conditions, two flow regimes can be distinguished. 
First, in case of broadband actuation with frequencies comprising the natural wake time scale, the convection of the jet structures enhances wake entrainment, shortens the length of the recirculating flow and increases drag.  
Secondly, at higher actuation frequencies, shear-layer deviation leads to \textit{fluidic boat-tailing} of the wake.
It additionally lowers its turbulent kinetic energy thus reducing the entrainment of momentum towards the recirculating flow.
The combination of both mechanisms produces a raise of the base pressure and reduces the drag of the model. 
Both actuation regimes are characterized by complementary velocity, pressure and drag measurements at several upstream conditions and control parameters.
By adding curved surfaces to deviate the jets by the Coanda effect, periodic actuation is reinforced and drag reductions of about $20\%$ are achieved. 
The unsteady Coanda blowing not only intensifies the flow deviation and the base pressure recovery but also preserves the unsteady high-frequency forcing effect on the turbulent field.
The present results encourage further development of fluidic control to improve the aerodynamics of road vehicles and provide a complementary insight into the relation between wake dynamics and drag.  

\end{abstract}


\section{Introduction}

Drag reduction of bluff bodies has become a major challenge for transport industry due to increasing need for reducing fuel consumption and carbon pollution. 
As an example, the aerodynamic drag of ground vehicles accounts for more than $50\,\%$ of the engine power consumption on a highway \citep{Hucho93}. 
In contrast to streamlined bodies, blunt geometries induce massively separated flow with low pressure wake.
The dynamics of these wakes has been extensively investigated along the years \citep{Ahmed84} and is still subject of recent research \citep{Grandemange13,Choi14}. 
It contains a low kinetic energy recirculating motion surrounded by convectively unstable free shear-layers.
The recirculating flow itself is absolutely unstable and produces self-sustained large-scale flow oscillations \citep{Huerre90}. 

\begin{figure}
				        \centering
                \includegraphics[scale=0.48]{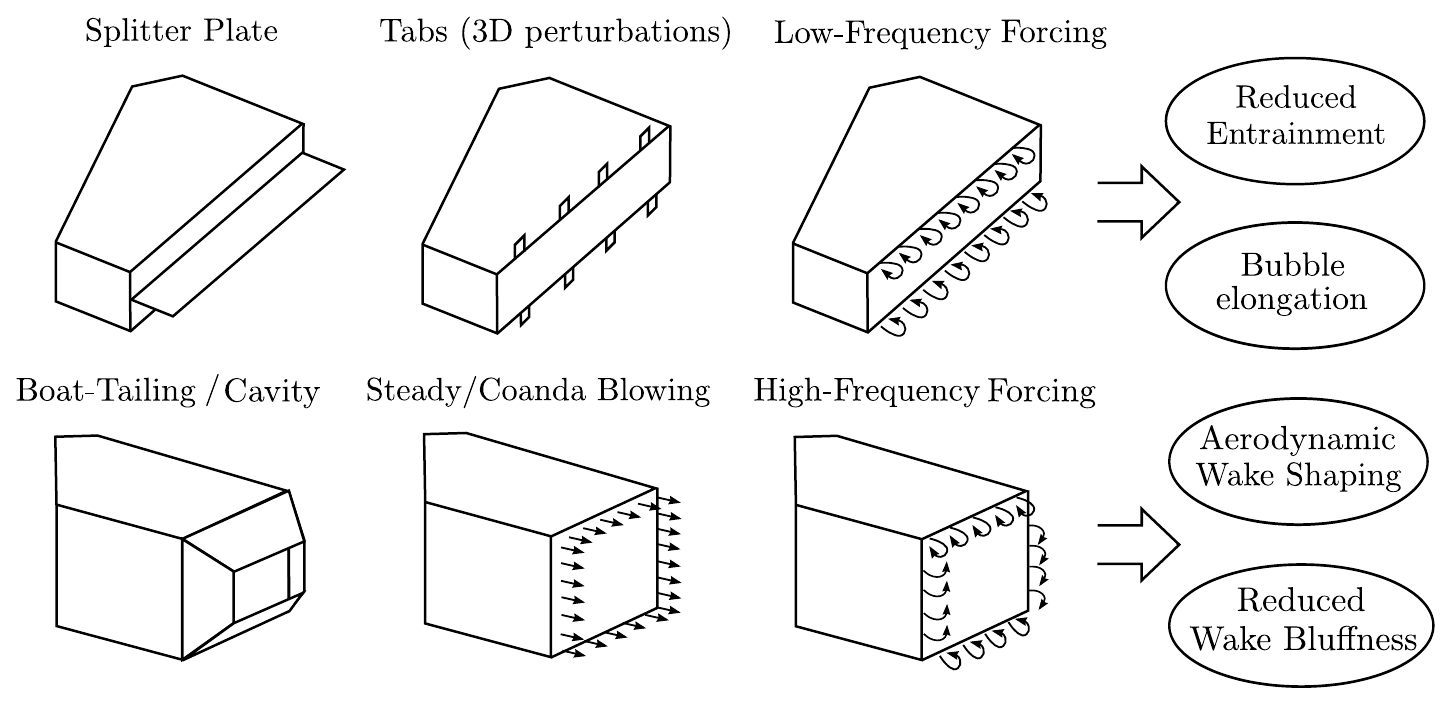}
                \caption{Overview of representative drag reducing devices for blunt body wakes.}
                \label{fig:fig1}
        \end{figure}

Within this context, flow control turned out to be an efficient way to modify bluff body wakes with the aim to increase the baseline pressure \citep{Choi08}.
A summary of the main drag reducing devices for blunt body wakes is drawn in figure~\ref{fig:fig1}.
The goal is either to reduce wake entrainment to elongate the formation region towards the Kirchhoff solution or to decrease the cross section of the wake in order to modify the bubble's aspect ratio and increase the pressure recovery \citep{Roshko55,Gerrard66}.  

A reduction of wake entrainment has been mainly achieved for nominally 2D flows dominated by vortex shedding. 
The damping of the strong periodic dynamics in these wakes is effective either by the use of splitter plates \citep{Bearman65} or by three-dimensional perturbations using tabs devices \citep{Park06}.
Moreover, fluidic control by pulsating and zero-net-mass-flux jets reveals positive effects on the model's drag \citep{Pastoor08}. 
However, the same strategies have not been successfully applied in 3D turbulent flows, where the vortex shedding amplitude is reduced.
In such wakes, pressure recovery is obtained by boat-tailing and cavity appendices \citep{Choi14,Evrard16}, or by producing flow deflection through steady jets associated or not with a Coanda effect \citep{Englar01,Littlewood12,Pfeiffer12}. 
For practical applications, such control techniques are limited with respect to power expense and geometrical constraints.

To find other alternatives, we explore here the possibility to further deviate the flow through unsteady forcing coupled with the Coanda effect.
The amplifier properties of the shear-layers allow to excite a wide range of vortical structures which may impact the wake dynamics \citep{Ho84, Fiedler98}. 

By local fluidic forcing with frequencies close to the shear-layer amplified modes, \citet{Chun96} drastically reduced the recirculating flow length behind a backward-facing step.
An increase of the actuation frequency, however, promoted an enlargement of the step's bubble.
Equivalent results have been obtained from the numerical simulations of \citet{Dandois07} and \citet{Dahan12} around ramps and steps. 
While low-frequency actuation amplifies the development of the shear-layers, high-frequency forcing stabilizes the velocity fluctuations lowering entrainment and increasing the bubble size or the base pressure. 
Although high-frequency forcing leads immediately after actuation to an increase of turbulent kinetic energy production, a noteworthy damping of flow unsteadiness is observed further downstream \citep{Vukasinovic10}.
This effect has been associated to a dissipative small-scale effect influencing notably the large-scale motions.
More recently, \citet{Parezanovic15} exemplified the capabilities of both increasing or decreasing the turbulent fluctuations across a canonical mixing layer by applying low or high-frequency forcing, respectively.

It is natural to extend such concepts to periodic forcing of 3D bluff body wakes for drag control.
Actuation at frequencies with an order of magnitude higher than the natural vortex shedding appears a promising strategy for wake manipulation since it allows to act directly on the spreading rate of the shear-layers, while the global instability modes of the flow are not amplified \citep{Glezer05}. 
Recently, \citet{Morrison09}, \citet{Barros14} and \citet{Oxlade15} applied small-scale actuation to reduce the pressure drag of axisymmetric and square-back geometries using unsteady jets.
They were periodically released along the border of the rear geometry in the direction of the main flow. 
A virtually shaped time-averaged flow was observed by these authors and associated to a reduction of the wake cross section and drag. 
However, the mechanisms by which the forced shear flow affects the base pressure remain to be clarified.   
One would also envisage coupling unsteady forcing and the Coanda effect to further deviate the flow.
Generally, defining how \textit{low} or \textit{high} is a driving frequency when compared to the wake vortex shedding is crucial to identify the physical mechanisms responsible for the drag changes in the light of their time scales \citep{Glezer05}.

The present work aims to shed further light on such aspects bringing out novel ways to manipulate bluff body drag.
For that, we apply periodic fluidic forcing along the trailing edges of a square-back geometry similar to that studied by \citet{Ahmed84}.
By varying both the excitation frequency and amplitude, our goal is to identify the effects of forcing on the wake and drag as well as to educe its physical mechanisms.
The experimental apparatus designed for this study is detailed in $\S\,\,2$, followed by a brief description of the unforced, reference flow in $\S\,\,3$.
A systematic study by varying the control parameters is presented in $\S\,\,4$, where we identify an increase and decrease of drag respectively for low and high-frequency actuation.
In $\S\,\,5$, we use velocity and pressure measurements to analyze how high-frequency forcing increases the base pressure.  
An extension of the actuation properties is given in $\S\,\,6$ by illustrating the effects of coupled unsteady Coanda blowing on the drag succeeded by our concluding remarks ($\S\,\,7$).

\section{Experimental configuration} 

This section describes the set-up of the bluff body arrangement inside the wind-tunnel working section and the different measurement techniques.
In addition, we expose details on the pneumatic actuation system which was specifically designed to influence the drag by modifications of the wake flow.
Complementary information on the entire apparatus can be found in \citet{Barros15a}. 

\subsection{Wind-tunnel facility and model geometry}

\begin{figure}
				        \centering
                \includegraphics[scale=0.38]{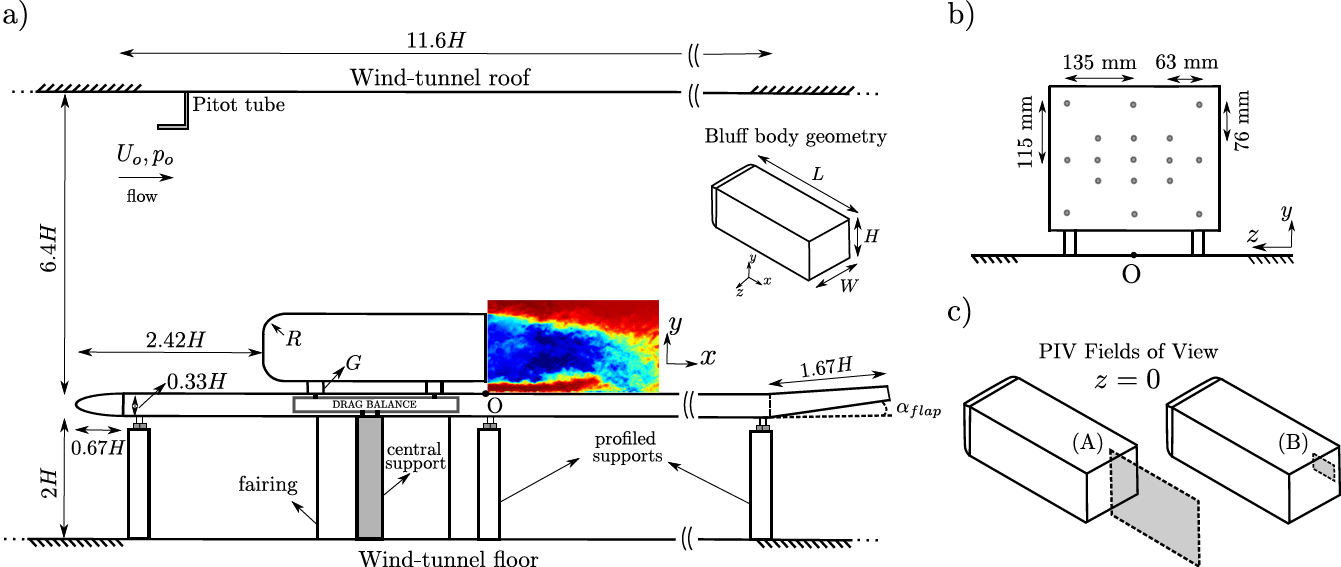}
                \caption{Experimental set-up. a) Wind-tunnel, flat plate dimensions and model positioning. The details of the model geometry are given by its 3D view in the inserted picture. b) Pressure taps locations used to evaluate the base pressure. c) PIV fields of view in the symmetry plane: a large window (A) to perform 2D2C PIV and a smaller window (B) for time-resolved PIV close to the upper shear-layer.}
                \label{fig:fig2}
        \end{figure}

The experiments are performed inside the working section of a subsonic wind-tunnel of $2.4\,\text{m}$ width and $2.6\,\text{m}$ height. 
The free-stream velocity $U_{o}$ varies between $10\,\text{m}\text{s}^{\text{-}1}$ and $20\,\text{m}\text{s}^{\text{-}1}$ with a turbulence intensity of the order of $0.5\,\%$ and no measurable velocity gradients upstream of the bluff body. 
Figure~\ref{fig:fig2}(a) displays a scheme of the bluff body arrangement inside the working section. 
The front edges of the bluff body are rounded with a radius $R=0.085\,\text{m}$.
The whole model with height $H = 0.297\,\text{m}$, width $W=0.350\,\text{m}$ and length $L=0.893\,\text{m}$ is fixed at a geometrical ground clearance of $G=0.05\,\text{m}$ above a flat plate floor.
The distance between the elliptical leading edge of this false floor and the model is $2.42H$.
The ground clearance $G$ is about five times greater than the oncoming boundary layer, which measures $\delta_{99\,\%}\sim0.034H$ with displacement thickness $\delta^\star\sim0.004H$ and shape factor $\overline{H}=1.58$.
To compensate the lift produced by the whole arrangement, a trailing edge flap inclined with angle $\alpha_{\text{flap}}=5.7\,^\circ$ allows to align the flow parallel to the false floor.
The overall blockage ratio of the model in the cross-section above the flat plate is $2.2\,\%$ making blockage corrections unnecessary. 

The Reynolds number based on the height of the model is defined as $Re_H = {U_o}H/{{\nu}}$, where $\nu$ is the kinematic viscosity of the air at ambient temperature. 
Most of the results are conducted for $U_{o}=15\,\text{m}\text{s}^{\text{-}1}$, corresponding to $Re_H = 3\times10^5$. 
For reference purposes, we introduce a Cartesian coordinate system with $x$ the streamwise, $y$ the transverse or cross-stream and  $z$ the spanwise directions. 
Its origin $O$ is arbitrarily set on the ground at the rear end of the bluff body.
Unless otherwise specified, all physical quantities are normalized by $U_{o}$, $H$ and by the dynamic pressure ${q_o}=0.5\rho{U_{o}}^2$, where $\rho$ is the air density at ambient temperature and pressure. 

\subsection{Drag measurements}

In order to quantify the effects of actuation on the drag, the model supports are directly fixed on an in-house one component aerodynamic balance with a high sensitive sensor (9217A Kistler piezoelectric, $1\,\text{mN}$ upwards).
The precision of the force measurements is $\pm0.07 \text{N}$. 
The measured drag force $F_x$ is normalized by the frontal area $S=WH$ of the bluff body as follows:

\begin{equation}
{C_x}= \frac{F_x}{{q_o}{S}},
\end{equation}
where $C_x$ is the drag coefficient.

The $C_x$ is calculated from six independent time averages which are all obtained during a single time interval of 10 seconds by the ETEP acquisition system.
The acquisition frequency is set to $6.25\,\text{kHz}$ and a low-pass filter is applied to obtain the mean drag coefficient. 
For example, at $Re_H=3.0\times{10^5}$ the drag coefficient of the baseline flow equals ${C_{x_{o}}}=0.293\pm0.005$ with $95\,\%$ confidence.
Based on the standard deviation from measurements with several configurations, the overall error in the drag coefficient is less than 2\%.
All drag measurements are corrected to take into account the thrust of the pulsed jets used for wake forcing.
This thrust is evaluated by measuring the reaction force caused by the expansion of the pulsed jets into quiescent air.
For the strongest jet blowing used in this study, this thrust achieves about 3-4\% of $F_x$ at $Re_H=3.0\times{10^5}$.

To quantify the drag variations with respect to the reference flow, we use the drag parameter ${\gamma}_{d}={{C_x}}/{C_{x_{o}}}$ which indicates an increase (resp. decrease) of drag for values greater (resp. smaller) than 1.

\subsection{Base pressure coefficient}

Pressure drag variations are measured by 17 pressure taps, which are distributed on the rear surface as illustrated in figure~\ref{fig:fig2}(b). 
As large eddy simulations of this bluff body configuration by \citet{Osth14} demonstrated, both the number and disposition of the pressure taps are sufficient to determine the base pressure distribution along the rear side of the model. 
The differential off-set pressure sensors (HCLA 02X5DB) operate in a range of $\pm\,250\,\text{Pa}$ with the upstream static pressure $p_o$ as reference.
The nominal response delay of these sensors is about $0.5\,\text{ms}$, which enables us to perform unsteady pressure measurements according to \citet{Ruiz09}. 
The acquisition frequency of all pressure taps is set to $6.25\,\text{kHz}$. 

By using the dynamic head $q_o$, the pressure coefficient reads:

\begin{equation}
 {C_p}= \frac{p-{p_o}}{{q_o}}.
\end{equation} 

We define $\langle{C_p}\rangle$ and $\overline{C_p}$ as the spatial and the time-averaged pressure coefficients respectively taken over the rear surface of the model and during 60 seconds of acquisition. 
Although the wake may exhibit long time scales associated to intermittent reversals of the recirculating flow \citep{Grandemange13}, the spatially and time-averaged pressure coefficient $\langle\overline{C_{p}}\rangle$ should not be affected according to the recent measurements from \citet{Volpe15}. 
 
Based on standard deviations from preliminary longtime acquisitions, the averaged base pressure for the unforced reference flow $\langle\overline{C_{p_{o}}}\rangle$ is measured with a precision of $\pm0.004$ corresponding to less than $3\,\%$ of $\langle\overline{C_{p}}\rangle$.
Like in the case of drag measurements, we define a pressure parameter ${\gamma}_{p}=\langle\overline{C_p}\rangle/\langle\overline{C_{p_{o}}}\rangle$: since the $C_p$ values along the base are negative, ${\gamma}_{p}\,<\,1$ (resp. ${\gamma}_{p}\,>\,1$) is associated with an increase (resp. decrease) of pressure.

\subsection{Velocity measurements in the wake}

Velocity measurements are performed in the wake by the use of particle image velocimetry (PIV) and hot-wire anemometry (HWA). 
Two PIV fields of view are located in the symmetry plane ($\text{z}=0$). 
The position of these fields is detailed in figure~\ref{fig:fig2}(c). 
A larger field (A) spans the whole wake containing entirely the recirculating flow domain. 
The second domain (B) corresponds to a zoom on the upper shear-layer extending to $x/H\sim0.4$. 

The largest field of view is used to capture the essential global modifications of the forced wake. 
Both the streamwise ($u$) and the cross-stream or transverse ($v$) velocity components of the flow are measured by two LaVision Imager pro X 4M cameras with resolution of $2000\times2000$ pixels. 
A laser sheet is pulsed with time delays of $120\,{\mu}\text{s}$ (when $U_{o}=15\,\text{m}\text{s}^{\text{-}1}$) in the symmetry plane and image pairs are acquired at a sampling frequency of $3.5\,\text{Hz}$. 
Velocity vectors are processed with an interrogation window of $32\times32$ pixels with a 50 \% overlap. 
The resulting spatial resolution is approximately $1\,\%$ of the model's height. 
Ensembles of 1000-2500 independent velocity fields are used to compute first and second order statistics. 

The second PIV set-up enables a zoom close to the upper edge of the model with a high-speed PIV (HSPIV). 
One Photron\textsuperscript{\textregistered} SA-Z 1.1 camera with a resolution of $1024\times1024$ pixels was used to acquire images at a rate of $10\,\text{kHz}$. 
The laser sheet is generated by a Quantronix MESA $532\,\text{nm}$ system. 
Most of the tests are performed during $8\,\text{s}$ with 80000 snapshots. 
The images are processed with an interrogation window of $16\times16$ pixels and an overlap of 50 \%.
The final spatial resolution is $0.003H$.

Hot-wire measurements using a single wire probe (55P11) are acquired by a StreamlinePro Anemometer System (from Dantec Dynamics\textsuperscript{\textregistered}). 
Typically, they are performed to quantify the boundary layer characteristics and to analyze the spectral content in the wake downstream the closure of the recirculating flow. 
The HWA also serves to calibrate the exit velocity of the pulsed jet system. 
A 55H21 support fixes the probe to a profiled displacement system installed on the roof of the wind-tunnel. 
The velocity measurements are sampled at $6.25\,\text{kHz}$ and the duration of each test varies between 60 and $120\,\text{s}$.

				\begin{figure}
				        \centering
                \includegraphics[scale=0.45]{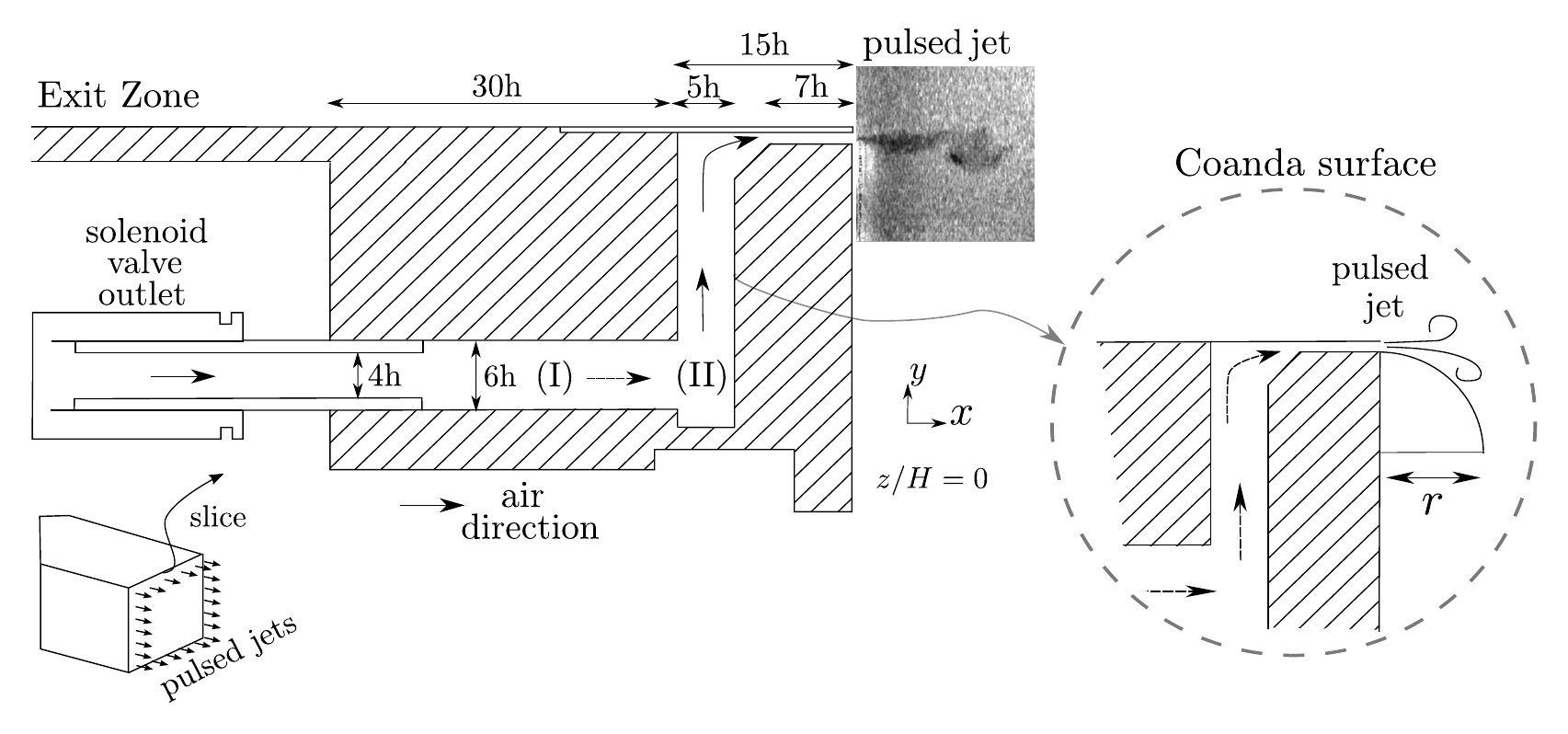}
                \caption{Sketch of the exit zone responsible for emission of the pulsed jets. It is located downstream the plenum chamber which feeds the solenoid valves with pressurized air. The periodic flow passes through an elbowed conduit until arriving at the exit slit with thickness $h=1\,\text{mm}$. An additional curved surface of radius $r=9\,\text{mm}$ can be installed to the set-up in order to produce a Coanda effect.}
                \label{fig:fig3}
        \end{figure}

\subsection{Pulsed jet system and forcing parameters}

As stated in the Introduction, wake forcing is obtained by unsteady, periodic jet blowing along the edges of the model \citep{Cattafesta11}.
The generation of pulsed jets with an exit velocity $V_j$ results from the pressure difference between the external flow and a pressurized air reservoir located inside the body.
The volume of this cylindrical plenum chamber is 3 liters and its internal pressure is called the input pressure $P_i$. 

The mass flow is driven periodically by 32 solenoid valves distributed homogeneously along the rear part of the model, upstream of the trailing edges, as illustrated in figure~\ref{fig:fig3} by the geometry details of the jet exit zone. 
The pulsating frequency $F_i$ is selected by a rectangular waveform with duty-cycle of $40\,\text{\%}$ and can be settled independently for each trailing edge, but all of them are connected to the same $P_i$. 
Although the number of solenoid valves is finite, the exit slit with thickness $h=1\pm0.1\,\text{mm}$ is continuous and spans all the periphery of the four edges.
The 32 outlets from the solenoid valves are followed by a circular pipe section (I in figure~\ref{fig:fig3}). 
All these pipes are equally spaced along the four trailing edges. 
In this conduit, the flow is convected along $x$ until arriving at the region (II), where the flow is free to diverge either along $y$ or $z$. 
The region (II) plays an important role to obtain an homogeneous outflow. 
Upstream of the exit slit, the flow passes through a sharp elbowed geometry.
This might introduce some additional vorticity in the boundary layers of the conduit, but an extension of length $7h$ helps to guide the flow up to the exit section.

				\begin{figure}
				        \centering
                \includegraphics[scale=0.32]{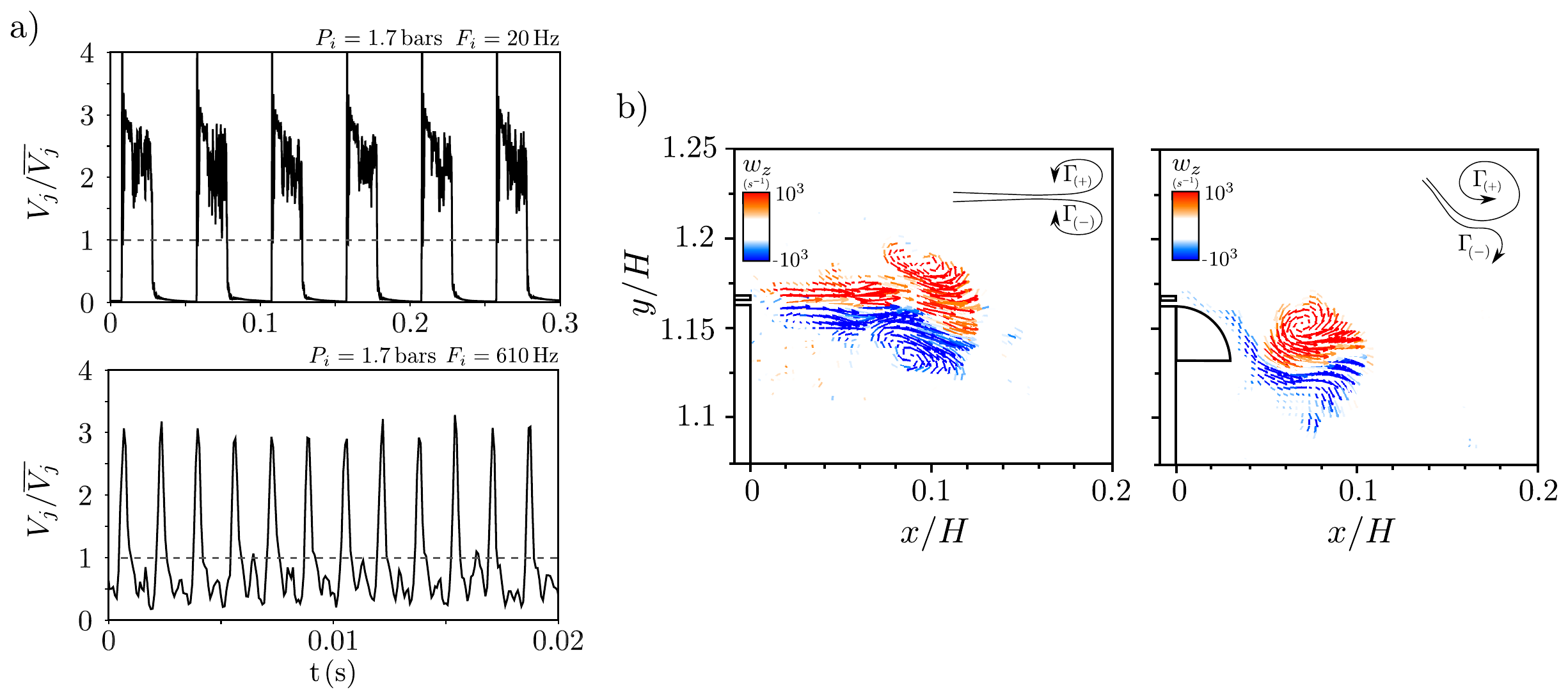}
                \caption{Pulsed jet velocity and vorticity field. a) Velocity signals $V_j$ for two configurations corresponding to an input pressure $P_i$ of $1.7\,\text{bars}$ and frequency $F_i$ of 20 and $610\,\text{Hz}$. b) Vorticity ($\omega_z$) field during the stroke phase of the pulsed jet in quiescent air at an actuation frequency of $F_i=20\,\text{Hz}$ corresponding to $St_H=0.4$ and $\overline{V_j}=6.5\,\text{m}\text{s}^{\text{-}1}$. The formation of counter rotating vortical structures is indicated by the signs of circulation $\Gamma$. The presence of the Coanda surface deviates the pulsed jets.}
                \label{fig:fig4}
        \end{figure}

Additionally, a Coanda surface can be added just bellow the jet exit as displayed in the same figure. 
This surface is a quarter of a disk and is installed along the four trailing edges. 
In the present study, the radius of this geometry is fixed to $r=9\,\text{mm}$, corresponding to $3\,\%$ of the model's height.
When the Coanda surface is installed, $\langle\overline{C_{p_{o}}}\rangle$ increases $3\,\%$ at $Re_H=3.0\times{10^5}$.
This is due to a slight passive flow deviation in view of the rounded surface \citep{Littlewood10} and the inherent cavity effect created by the added geometry \citep{Evrard16}.

The jet velocity is quantified by a single hot-wire probe $1\,\text{mm}$ downstream and at the centerline of the exit slit. 
The effective jet velocity $V_{\text{eff.}}$ and the momentum coefficient $C_\mu$ of the jet read:

\begin{equation}
{V_{j_{\text{eff.}}}}=\sqrt{{\overline{V_j}}^2+{({V_j}_{\text{Sdv.}})}^2},
\end{equation}  

\begin{equation}
C_\mu=\frac{{s_j}\overline{{V_j}^2}}{{S}{{U_o}}^2}=\frac{{s_j}{(V_{j_{\text{eff.}}}})^2}{{S}{{U_o}}^2},
\end{equation}
where $\overline{V_j}$, ${V_j}_{\text{Sdv.}}$ and $s_j$ are the time-averaged jet velocity, its standard deviation and the jet slit cross-sectional area, respectively. 
Examples of velocity signals for two selected frequencies are reported in figure~\ref{fig:fig4}(a).

The effective jet velocity ${V_{j_{\text{eff.}}}}$ takes into account not only the steady component of the jet velocity but also its oscillatory dynamics. 
Although the exit slit is 2D due to the designed continuous geometry, the jet velocity presents spanwise variations due to the complex upstream flow conditioning.
Measurements of ${V_{j_{\text{eff.}}}}$ along the spanwise direction indicate variations of at most $\pm\,10\,\%$ compared to the spanwise averaged velocity.

The quantities $\overline{V_j}$ and ${V_{j_{\text{eff.}}}}$ are calculated from time-averaging during a large number of pulsation cycles and take into account the time when there is no discharge of fluid from the exit slit. 
Details about the pulsed jet formation are further characterized by the non-dimensional stroke length ${L_o}=\frac{1}{h}\int_0^\tau {V_j}(t)dt$ as defined in \citet{Glezer02}.
More specifically, it is possible to show that ${L_o}={\overline{V_j}}{T_i}/h$ with the pulsating period $T_i={F_i}^{-1}$ \citep{Smith98}.
As an example, ${L_o}=289\pm3$ and ${L_o}=14.1\pm0.2$ for the velocity signals displayed in figure~\ref{fig:fig4}(a) respectively when ${F_i}=20\,\text{Hz}$ and ${F_i}=610\,\text{Hz}$.  

Finally, for spectral analysis, we consider the actuation frequencies in non dimensional form by defining a Strouhal number based on the model's geometry $St_H={{F_i}H}/{{U_o}}$. 
The pulsed jet vorticity field $\omega_z$ during the stroke phase is exemplified in figure~\ref{fig:fig4}(b) with and without the Coanda surface. 
The actuation parameters are $St_H=0.4$ and $\overline{V_j}=6.5\,\text{m}\text{s}^{\text{-}1}$ corresponding to ${L_o}=325\pm3$.

\section{Baseline flow} 

Before analyzing the effects of actuation on the bluff body drag, we briefly review some aspects of its unforced flow. 
The statistical properties of the recirculating region as well as specific global quantities in the wake are presented.
We also quantify the unsteady shedding mode encountered in this flow.

\begin{figure}
				        \centering
                \includegraphics[scale=0.39]{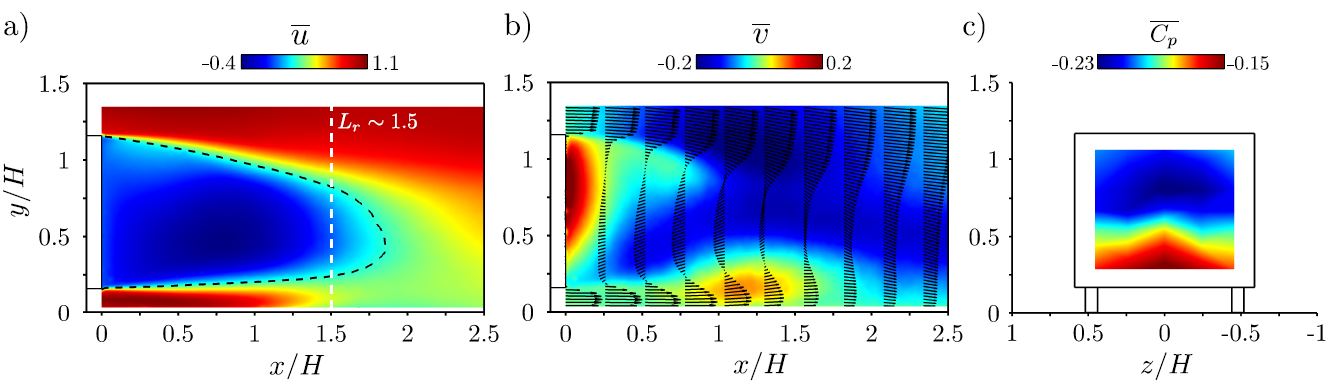}
                \caption{Mean properties of the baseline wake flow at ${Re_H}=3\times10^{5}$. a) Time-averaged streamwise velocity $\overline{u}$ with the iso-value line $\overline{u}=0.25$. The recirculating flow length $L_r$ is displayed by the vertical dashed line. b) Velocity vectors and cross-stream velocity ($\overline{v}$) distribution. c) Mean base pressure coefficient on the rear surface.}
                \label{fig:fig5}
        \end{figure}

The time-averaged velocity components and base pressure distribution are presented in figure~\ref{fig:fig5} at ${Re_H}=3\times10^{5}$.
The negative values of $\overline{u}$ in figure~\ref{fig:fig5}(a) indicate the existence of a recirculating flow field.
Its shape is suggested by the displayed isoline $\overline{u}=0.25$.
The length of the recirculating flow zone is defined by:

\begin{equation}
 {L_r}= \max(x/H)_{\overline{u}(x/H)\leq0},
\end{equation}
from which one can estimate the apparent mean bubble length.
For the present Reynolds number, the measured $L_r=1.50\pm0.01$ agrees with other numerical or experimental studies \citep{Wassen10,Lahaye14,Volpe15} of this flow configuration.
 
From the $\overline{v}$ distribution in figure~\ref{fig:fig5}(b), we conclude that the mean wake is dominated by a large clock-wise recirculating motion.
This transverse flow asymmetry in the central middle plan is expected due to the presence of the ground, which not only affects the developing boundary layers along the model but also modifies the underflow boundary conditions\footnote{It is worth to emphasize here that our proceeding results on wake forcing remains unchanged whether the flow asymmetry appears along the top/bottom or the lateral directions \citep{Barros15a}, hence validating our analysis in the symmetry plane of the configuration. However, attention must be paid to the inherent end-effects of the bluff body which are not present in our in-plane analysis.}.

This is particularly clear when analyzing the base pressure distribution in figure~\ref{fig:fig5}(c). 
Close to the upper edge, a low-pressure zone is established where ${(\overline{C_{p}})}_{\text{min}}\sim-0.23$.
It is associated with the large clock-wise recirculating motion which considerably curves the flow streamlines generating high-pressure gradients and a local pressure drop.
Along the lower edge of the model, the pressure is higher ${(\overline{C_{p}})}_{\text{max}}\sim-0.14$.  
Globally, the spatially averaged base pressure is $\langle\overline{C_{p_{o}}}\rangle=-0.201$, comparing well with values behind this geometry of \citet{Krajnovic03} and \citet{Grandemange13}.

In table~\ref{tab:table1}, we summarize the time-averaged quantities for varying Reynolds numbers corresponding to $U_{o}=10,15\,\text{and}\,20\,\text{m}\text{s}^{\text{-}1}$, where we additionally include the mean drag coefficient ${C_{x_{o}}}$.
There is a decrease of ${C_{x_{o}}}$ from 0.31 to about 0.27 when increasing the Reynolds number, with a simultaneous augmentation of $\langle\overline{C_{p_{o}}}\rangle$ and $L_r$.
This tendency has been also observed by \citet{Lahaye14} and \citet{Volpe15} among others and might be associated to the boundary layer characteristics prior to separation.
Indeed, \citet{Spohn02} showed the presence of flow detachment on the front curved edges of the model, whose convected perturbations may impact the separating boundary layer at the rear surface of the model and the wake.
The resulting effect is a significant scatter of the drag values found in the literature, in which ${C_{x_{o}}}$ ranges from 0.25 to up 0.36 \citep{Ahmed84,Krajnovic03,Wassen10,Grandemange13,Lahaye14}.

\begin{table}
  \begin{center}
   \begin{tabular}{cccccc}
    Reynolds number - ${{Re}_H}$  & ${L_r}$   & $\langle\overline{C_{p_{o}}}\rangle$  & ${(\overline{C_{p}})}_{\text{min}}$ & $(\overline{C_{p}})_{\text{max}}$  & ${C_{x_{o}}}$ \\
       $2\times{10^5}$  & $1.45\pm0.01$  & -0.204 &-0.240 & -0.139 & $0.308\pm0.005$    \\
       $3\times{10^5}$  & $1.50\pm0.01$  & -0.201 &-0.233 & -0.144 & $0.293\pm0.005$  \\
       $4\times{10^5}$  & $1.55\pm0.01$  & -0.196 & -0.219& -0.157 & $0.269\pm0.005$   \\
  \end{tabular}
  \caption{Recirculation length, base pressure and drag coefficients for varying $Re_H$ number. The precision on the $C_p$ values is $\pm0.004$.}
  \label{tab:table1}
  \end{center}
\end{table}

The turbulent conditions of the wake are quantified by measuring the boundary layers prior to separation.
The characteristics of the streamwise velocity profiles measured at the center of the upper and lateral edges are summarized for different Reynolds numbers in the table~\ref{tab:table2}.

The calculated shape factors $\overline{H}=\delta^{\ast}/{\theta}$, where $\delta^{\ast}$ and ${\theta}$ are respectively the displacement and momentum deficit thickness, are considerably lower than the reference Blasius solution value of 2.59.
Furthermore, by estimating the friction velocity $u_{\tau}$, the velocity profiles exhibit a log-law behavior which is further detailed in \citet{Barros15a} and not presented here for brevity.

\begin{table}
\begin{center}
\begin{tabular}{ccccccc}

${{Re}_H}$    & $\delta_{99}$ & $\delta^{\ast}$ & $\theta$  & $\overline{H}$ & $u_{\tau}$ \\

$2\times{10^5}(\text{upper})$         & 0.33 & 0.038 & 0.030 & 1.27 & 0.020      \\
$3\times{10^5}(\text{upper})$         & 0.22 & 0.026 & 0.020 & 1.33 & 0.025      \\
$4\times{10^5}(\text{upper})$         & 0.09 & 0.014 & 0.010 & 1.38 & 0.041      \\
$2\times{10^5}(\text{lateral})$         & 0.18 & 0.024 & 0.015 & 1.63 & 0.028      \\
$3\times{10^5}(\text{lateral})$         & 0.12 & 0.017 & 0.010 & 1.70 & 0.033      \\
$4\times{10^5}(\text{lateral})$         & 0.09 & 0.014 & 0.008 & 1.61 & 0.041      \\
\end{tabular}
\caption{Boundary layer characteristics at the upper and lateral trailing-edges.}
\label{tab:table2}
\end{center}
\end{table}

\begin{figure}
				        \centering
                \includegraphics[scale=0.35]{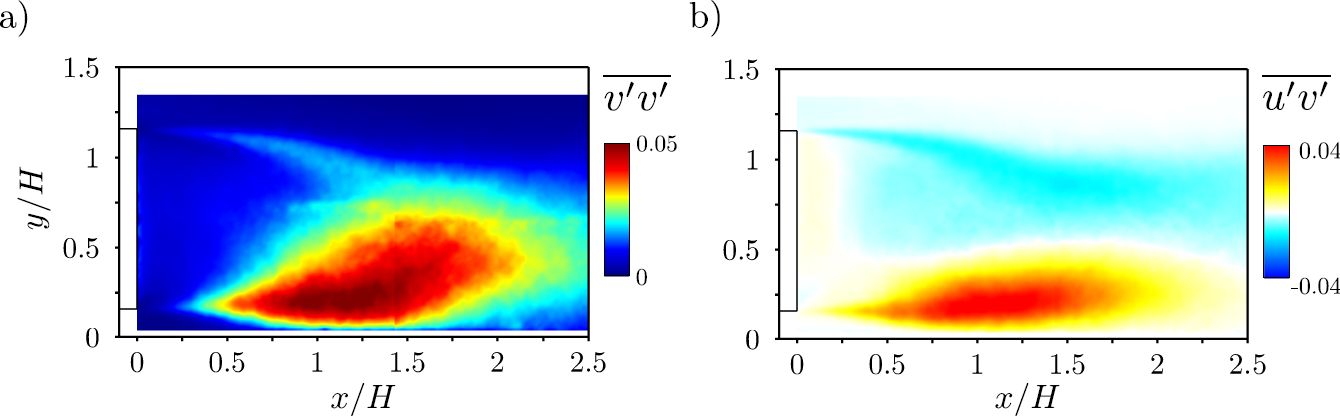}
                \caption{Second order statistical moments of the mean flow. a) Cross-stream velocity variance $\overline{v'v'}$. b) Streamwise-transverse velocity covariance $\overline{u'v'}$.}
                \label{fig:fig6}
        \end{figure}

The spatial signature of the shear-layer instabilities along the separated flow is described in figure~\ref{fig:fig6} by the velocity correlations $\overline{v'v'}$ and $\overline{u'v'}$.
As previously discussed, the ground proximity considerably impacts the distribution of the cross-stream velocity variance.
There is a concentration of flow unsteadiness along the lower shear-layer near the ground.
The velocity covariance $\overline{u'v'}$ better represents the spatial development of both shear flows up to the closing of the bubble.
Extreme values of these quantities together with the streamwise variance $\overline{u'u'}$ are reported in the table~\ref{tab:table3}.  

 \begin{table}
  \begin{center}
   \begin{tabular}{ccccccc}
    ${{Re}_H}$  & ${(\overline{u'v'})}_{\text{min}}$ & ${(\overline{u'v'})}_{\text{max}}$  & $\overline{u'u'}$ & $\overline{v'v'}$ & $(f_{n_o})_y$ & $(f_{n_o})_z$ \\
       $2.{10^5}$  &$-0.016$& $0.048$ & $0.101$ & $0.055$&$0.215$& $0.179$    \\
       $3.{10^5}$  &$-0.015$& $0.044$ & $0.095$ & $0.053$&$0.205$& $0.168$    \\
       $4.{10^5}$  &$-0.017$& $0.040$ & $0.090$ & $0.047$&$0.201$& $0.161$    \\
  \end{tabular}
  \caption{Extreme values of the time-averaged velocity fluctuations and oscillatory wake modes for varying $Re_H$ number. The precision of the oscillatory mode frequency is $\pm 0.011$}
  \label{tab:table3}
  \end{center}
\end{table}	
				
We additionally report the dimensionless frequency ${f_{n_o}}={f_{\text{VS}}H/{U_o}}$, where $f_{\text{VS}}$ is the vortex shedding frequency obtained from HWA spectra.
According to \citet{Grandemange13}, two oscillatory time-scales ${(f_{n_o})_y}$ and ${(f_{n_o})_z}$ are associated respectively to the the top/bottom ($y$) and lateral ($z$) interaction of the shear-layers.
In the present work, they are measured at two locations: in the symmetry plane of the configuration at $(x/H,y/H,z/H)=(2.0,0.9,0)$ and in a horizontal plane at $(2.0,0.6,0.5)$.
The values of ${(f_{n_o})_y}$ are close to 0.2, typically found for axisymmetric bluff bodies \citep{Rigas14}.
The lateral shedding values are of the order of 0.17 but, when normalized by $W$ and not $H$, agrees well with the 0.2 value.

\section{Overview of periodic forcing effects on the drag} 

We describe now the impact of periodic blowing on the baseline pressure and drag.
Systematic variations of the actuation frequency and amplitude as well as the influence of the upstream flow velocity are considered.

\subsection{Global effects of actuation} 

\begin{figure}
				        \centering
                \includegraphics[scale=0.8]{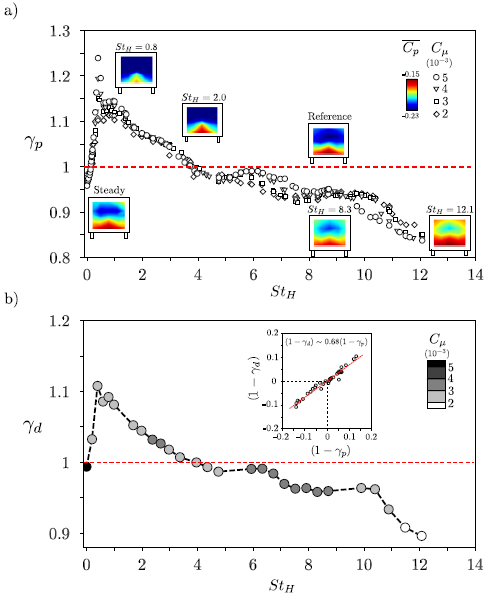}
                \caption{Base pressure and drag evolution as a function of the actuation parameters. a) Impact of $St_H$ and $C_\mu=\{2,3,4,5\}\times{10^{-3}}$ on $\gamma_p$. Changes of the $\overline{C_p}$ on the rear surface are indicated for representative configurations. b) Drag parameter $\gamma_d$ changes for varying actuation frequencies. The drag measurements are performed at a fixed input pressure $P_i$ of $1.45\,\text{bar}$. The color bar indicates the corresponding values of momentum coefficient for each measured point of the curve. An inserted correlation plot between $\gamma_d$ and $\gamma_p$ indicates that the drag variations correspond to 68\% of the base pressure changes.}
                \label{fig:fig7}
        \end{figure}

Unless otherwise stated, base pressure changes ($\gamma_p$) are quantified by maintaining the same actuation frequency $St_H$ along all edges.
The global impact of forcing on $\gamma_p$ is summarized in figure~\ref{fig:fig7}(a) for several values of $C_\mu$ at a Reynolds number $Re_H=3\times{10^5}$.

For very low frequencies corresponding to $St_H\in[0,0.1]$, the quasi-steady jet recovers up to $5\,\%$ of the base pressure, comparing well with past numerical \citep{Roumeas09,Wassen10} and experimental \citep{Krentel10,Littlewood12} results on steady blowing with similar exit jet velocities ${V_j}$.

This tendency is inverted at higher driving frequencies for $St_H\in[0.1,3.7]$, where the base pressure decreases instead.
Two different mechanisms contribute to this rapid increase of drag. 
Firstly, an abrupt augmentation of $\gamma_p$ at $St_H\sim0.4$ is related to the synchronization of actuation vortices with the oscillatory wake mode $(f_{n_o})_y$ decreasing the base pressure by up to $25\,\%$ when ${C_\mu}=5\times10^{-3}$.
At the same time, a significant shortening of the wake length is observed \citep{Barros16}.
Secondly, a range of actuation frequencies around $St_H\sim0.8$ provokes an increase of $\gamma_p$ leading to maximum amplitudes of 1.15 at $St_H\sim0.8$ and $C_\mu=5\times{10^{-3}}$. 
Overall, an increase of the jet amplitude $C_\mu$ leads to a drop of base pressure close to $St_H\sim0.8$.

A further rise of $St_H$ beyond 4.0 enables a gradual base pressure recovery.
The maximum 16\% raise of $\langle\overline{C_{p}}\rangle$ is achieved by the highest actuation frequency ($St_H=12.1$).
Interestingly, the numerical simulations from \citet{Dahan12} illustrated an increase of the step base pressure at high-frequency forcing, in agreement with recent tests conducted on axisymmetric wakes \citep{Oxlade15}.

Corresponding drag measurements at different actuation frequencies and amplitudes confirm this trend.
The corresponding evolution of $\gamma_d$ is displayed in figure~\ref{fig:fig7}(b).
For experimental simplicity, the input pressure is set to $P_i=1.45\,\text{bar}$, thus leading to several levels of $C_\mu$ as a function of the actuation frequency.
This should not invalidate our analysis since the impact of $C_\mu$ on $\gamma_p$ appears secondary compared to the influence of the actuation frequency $St_H$ at fixed $Re_H$ value. 
By comparing figures~\ref{fig:fig7}(a,b), we remark a strong connection between the evolution of $\gamma_p$ and $\gamma_d$. 
Drag alterations of $\pm10\,\%$ are measured at this Reynolds number.
The correlation between these quantities is confirmed by a linear fit displayed in the inserted picture, whose coefficient 
reveals that $68\,\%$ of base pressure changes are converted into drag modifications, in excellent agreement with the numerical findings from \citet{Krajnovic03}.

\subsection{Scaling of drag changes} 

\begin{figure}
				        \centering
                \includegraphics[scale=0.45]{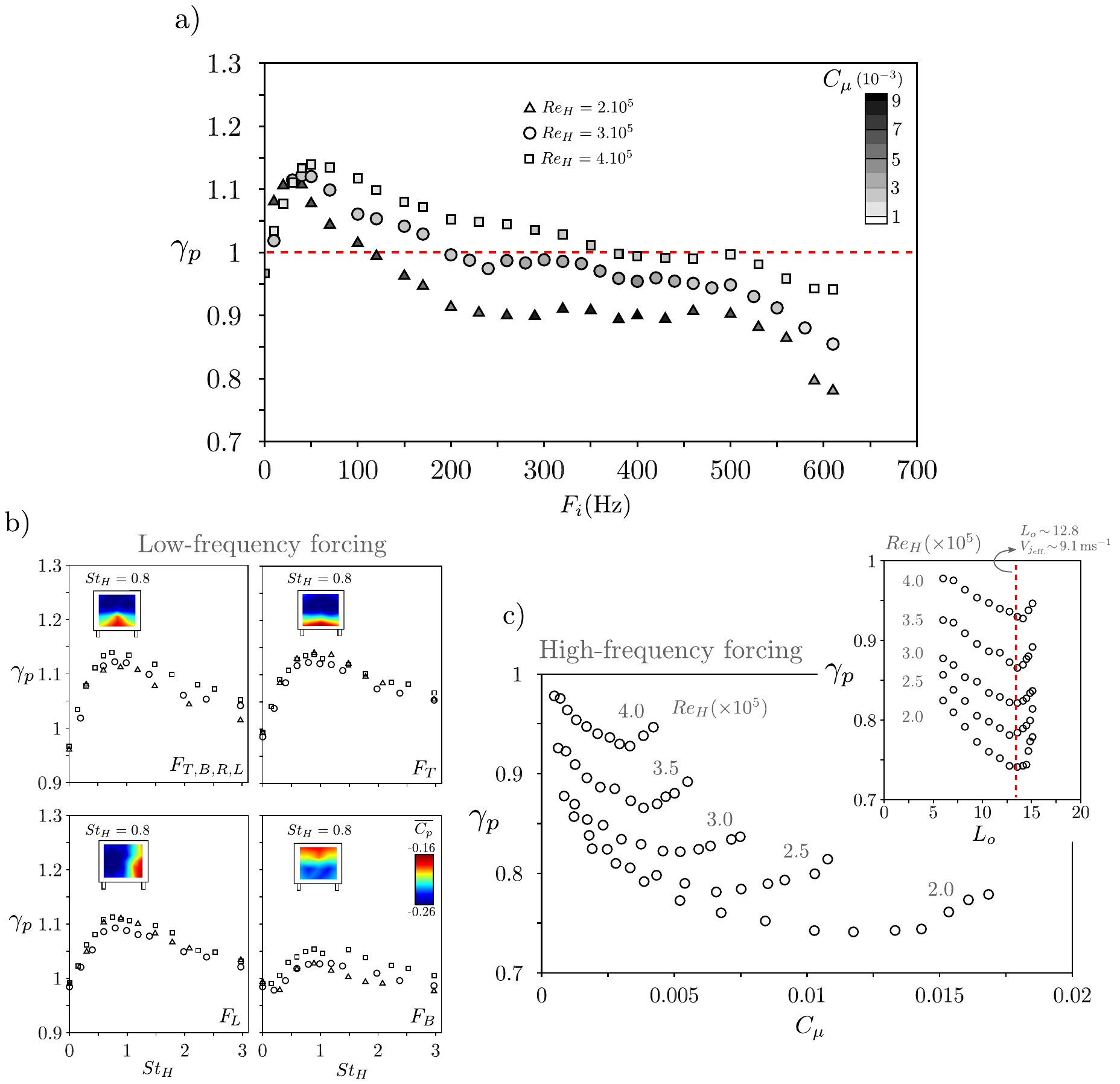}
                \caption{Scaling of drag changes. a) Reynolds-number effect on the base pressure changes for variable forcing frequency $F_i$. b) Scaling of the low-frequency forcing with constant jet amplitude $V_{j_{\text{eff.}}}\sim7.3\,\text{m}\text{s}^{\text{-}1}$ for all slit ($F_{T,B,R,L}$) and single-slit ($F_T$, $F_B$ and $F_L)$ actuation. The base pressure distribution of $\overline{C_p}$ for $St_H\sim0.8$ is displayed in the inserts. c) Effects of $Re_H$ and jet amplitude on the base pressure recovery at high-frequency forcing $F_i=610\,\text{Hz}$ along all slits. The studied Reynolds numbers ${Re_{H}}=\{2.0, 2.5, 3.0, 3.5, 4.0\} \times 10^5$ are indicated in the graph. The jet amplitude is described by the stroke length $L_o$ and the vertical dashed line indicates its optimum value $L_o\sim12.8$, corresponding to an effective jet velocity $V_{j_{\text{eff.}}}\sim9.1\,\text{m}\text{s}^{\text{-}1}$ for all upstream conditions.}
                \label{fig:fig8}
        \end{figure}

To gain further insight on the scaling of the drag changes, we present in figure~\ref{fig:fig8}(a) the effects of $F_i$ on the base pressure at three values of $Re_H=\{2,3,4\}\times{10^5}$, when actuation is applied along the four edges.
Clearly, two domains can be distinguished depending on whether $\gamma_p$ varies with both $F_i$ and $Re_H$.
First, an increase of $Re_H$ enlarges the low-frequency range in which the base pressure decreases ($\gamma_p>1$).
However, at actuation frequencies higher than $400\,\text{Hz}$, the evolution of $\gamma_p$ with forcing frequency remains similar for all Reynolds numbers, despite the different magnitudes of base pressure and forcing amplitude ($C_\mu$).

In the \textit{low-frequency} range, when actuating with a constant jet amplitude ($V_{j_{\text{eff.}}}\sim7.3\,\text{m}\text{s}^{\text{-}1}$ is displayed here), $\gamma_p$ scales with $St_H$, as shown by the collapsed data of figure~\ref{fig:fig8}(b). 
This scaling remains the same whether the forcing is applied along all edges ($F_{T,B,R,L}$) or only along individual slits $F_T$ (top), $F_B$ (bottom) and $F_L$ (left).

In Appendix A, we show that forcing at $St_H\sim0.8$ leads to an increase of the fluctuating velocities along the full recirculating flow.
One may therefore wonder if the increase of drag at this actuation frequency is associated to the amplification of instabilities along the shear-layers from separation.
Most amplified frequencies are usually scaled with the momentum thickness of the boundary layer, such that $St_\theta=F_i\theta/U_o$, in the range of $St_\theta=0.022-0.0024$ for turbulent shear-layers \citep{Ho84,Morris03}.
Previous results, however, illustrate strong scatter related to such a critical $St_\theta$ for which the recirculating flow region appears significantly shortened and the drag is increased.
For example, experiments conducted on step flows reported frequencies of about $St_\theta\sim0.01$ ($St_H\sim0.27$, based on the step height) \citep{Chun96,Dahan12,Garrido15} while actuation in wake flows present critical values of $St_\theta\sim0.02$ ($St_H\sim1.8$) for axisymmetric geometries \citep{Morrison09,Oxlade15} and $St_\theta\sim0.003-0.005$ ($St_H\sim0.12-0.8$) for nominally 2D or 3D square-back models \citep{Chaligne13,Chaligne13b}.
In the present work, $St_H\sim0.8$ scales the increase of drag for varying $Re_H$, but, when $St_\theta$ is calculated, we do not obtain a similar critical scaling for each upstream velocity $U_o$, suggesting that $\theta$ is not the right length scale of the problem.
Based on the seminal paper from \citet{Oster82}, however, we could conjecture that the forcing time scales are related to the most amplified frequencies further downstream, and this aspect is discussed in Appendix A.

An amplification of the vortices emitted by actuation along the shear-layer is observed in all studies mentioned above. 
Consequently, the circulation introduced at each jet slit may induce higher entrainment rates and a change of the mean large-scale recirculating flow orientation, evidenced here by the $\overline{C_p}$ distribution for each configuration (inserted pictures).
Additional information is provided and discussed in Appendix A.
 
These results also underline the inadequacy of $C_\mu$ to describe the effect of the forcing amplitude in the low-frequency range.
Indeed, despite $C_\mu\propto{s_j}{U_o}^{-2}$ at constant $V_{j_{\text{eff.}}}$, variations of $s_j$ (by reducing the number of actuated slits from four to one) or $Re_H$ (by changing $U_o$) do not impact $\gamma_p$ significantly.
On the other hand, variation of $V_{j_{\text{eff.}}}$ at a constant $Re_H$ implies changes of $\gamma_p$ \citep{Barros16}, indicating that the injected pulsed jet vorticity governs the jet amplitude scaling, whose physical details are beyond the scope of the present paper and needs to be further clarified by systematic measurements.   

The decrease of drag by \textit{high-frequency} forcing critically depends on the jet amplitude.
In figure~\ref{fig:fig8}(c), we display the impact of $C_\mu$ when actuation is performed at a forcing frequency $F_i=610\,\text{Hz}$ ($St_H=12.1$ at ${Re_{H}}=3.0\times 10^5$) for variable Reynolds number.
Base pressure parameter $\gamma_p$ drops monotonically by increasing $C_\mu$ and decreasing $Re_H$ up to an optimum.
This optimality has been similarly observed in the recent axisymmetric wake measurements from \citet{Oxlade15} at a fixed $Re_H$ value.
Our results show that the optimum decrease of drag does not scale only with $C_\mu$. 
Rather, it is measured at a constant stroke length $L_o\sim12.8$ corresponding to an effective jet velocity $V_{j_{\text{eff.}}}\sim9.1\,\text{m}\text{s}^{\text{-}1}$ for all Reynolds numbers ${Re_{H}}=\{2.0, 2.5, 3.0, 3.5, 4.0\} \times 10^5$, indicating that the intrinsic jet dynamics governs the forcing amplitude together with $C_\mu$ at high-frequency forcing.

\section{Drag reduction by high-frequency actuation}		   

In this section, we focus on the physical mechanisms behind drag reduction at actuation frequencies one order of magnitude higher than the vortex shedding (VS) mode and decoupled to its dynamics \citep{Glezer05}.
We consider here the previous high-frequency forcing frequency $F_i=610\,\text{Hz}$ selected along all trailing edges.
How does the actuation impact the base pressure and how do the wake changes depend on the upstream flow and jet amplitude?
We address these two questions to propose a conceptual scenario of forcing at this time scale.  
		
\subsection{Overview of the main effects} 		   

\begin{figure}
				        \centering
                \includegraphics[scale=0.5]{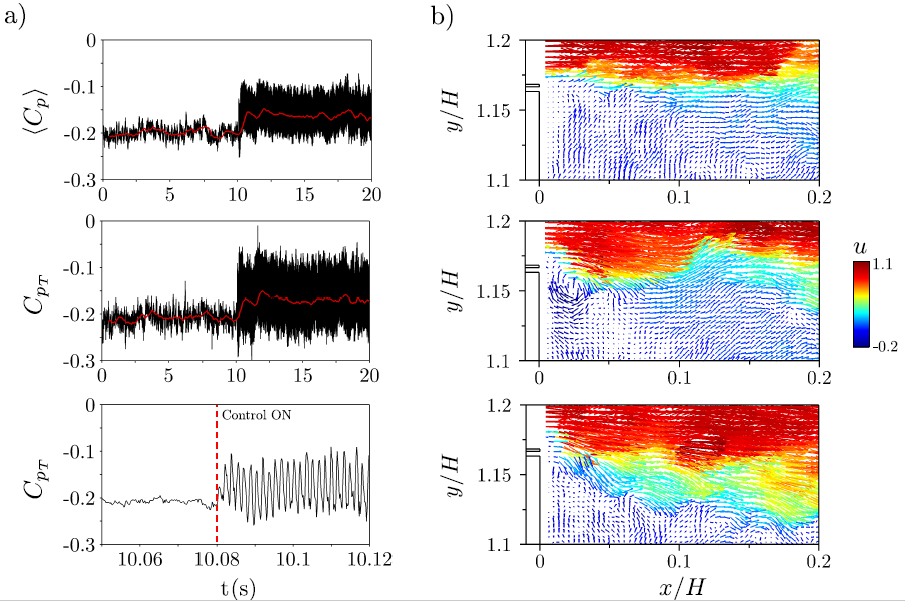}
                \caption{Transient response of high-frequency control applied with $St_H=12.1$ ($L_o\sim12.8$ and $V_{j_{\text{eff.}}}\sim9.1\,\text{m}\text{s}^{\text{-}1}$) at ${Re_{H}}=3 \times 10^5$. a) Time evolution of the base pressure $\langle{C_{p}}\rangle$ (top) and the pressure coefficient $C_{p_T}$ (middle and bottom). b) Snapshots of the velocity field: unforced flow prior to actuation (top), during actuation stroke (middle) and shear-layer vectoring (bottom). The vectors are colored by the streamwise velocity $u$.}
                \label{fig:fig9}
        \end{figure}

We proceed by investigating the transient response when actuation is set at $St_H=12.1$ and applied with jet amplitude $C_\mu=4.6 \times 10^{-3}$ ($L_o\sim12.8$ and $V_{j_{\text{eff.}}}\sim9.1\,\text{m}\text{s}^{\text{-}1}$) at ${Re_{H}}=3 \times 10^5$.
It is shown in figure~\ref{fig:fig9}(a) that the control suddenly increases $\langle{C_p}\rangle$ and the top pressure coefficient $C_{p_T}$ located at ($y/H\!=\!1.05$, $z\!=\!0$), as indicated by their filtered evolution represented by the red lines.
A magnified view in the bottom figure details the pressure oscillations associated to the pulsed jet frequency.
The pressure magnitude alternates between the unforced and higher values, leading to a $18\,\%$ increase of $\langle{\overline{{C_p}}}\rangle$.

Figure~\ref{fig:fig9}(b) displays three snapshots of the velocity field during this transition.
The unforced shear-layer prior to actuation is depicted in the top.
A circulatory flow is formed very close to the edge when starting control as shown in the middle picture.
The injected circulation induces a notable deviation of the shear-layer, quite similarly to the flow vectoring effect reported in \citet{Smith02} using adjacent synthetic-jet forcing.

\begin{figure}
				        \centering
                \includegraphics[scale=0.25]{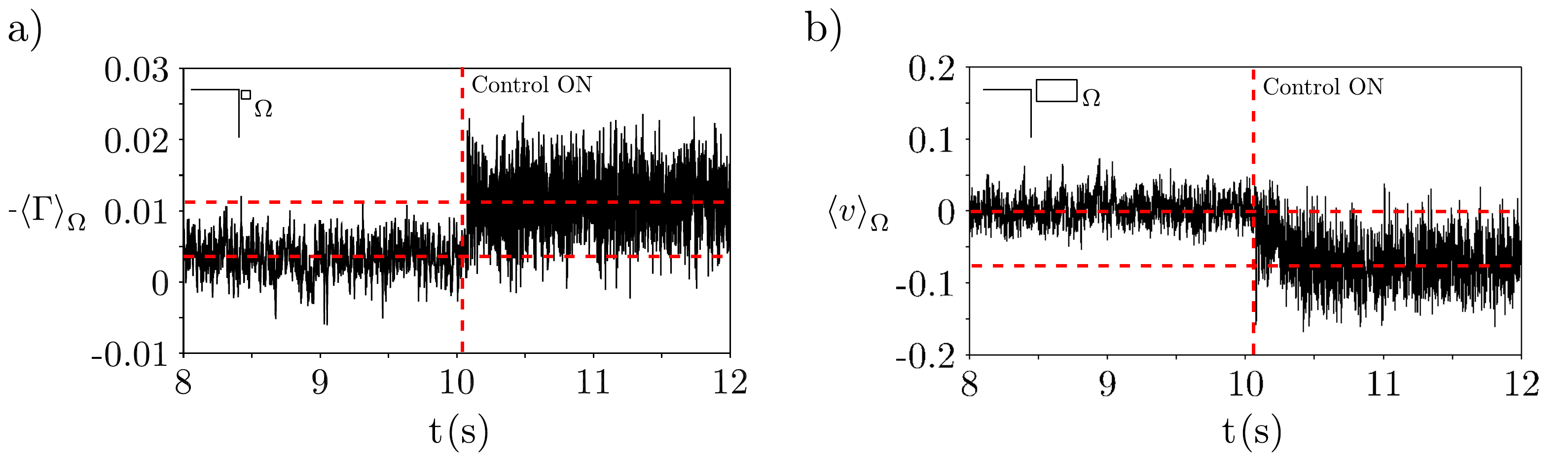}
                \caption{Effects of forcing on the near-edge circulation and the shear-layer vectoring at ${Re_{H}}=3 \times 10^5$. a) Circulation ${\langle{\Gamma}\rangle}_{\Omega}$ in the domain $\Omega$ limited by $(x/H,y/H)=([0,0.03],[1.14,1.17])$ adjacent to the exit jet slit. b) Time response of the integrated cross-stream velocity ${\langle{v}\rangle}_{\Omega}$ over the rectangular domain $\Omega$ limited by $(x/H,y/H)=([0,0.07],[1.15,1.19])$.}
                \label{fig:fig10}
         \end{figure}

In figure~\ref{fig:fig10}, we quantify this process by computing the circulation and the cross-stream velocity in the vicinity of the jet slit.
The integration of the vorticity field inside a rectangular region $\Omega$ limited by $(x/H,y/H)=([0,0.03],[1.14,1.17])$ shows an increase of flow circulation ${\langle{\Gamma}\rangle}_{\Omega}$ inside this domain (see figure~\ref{fig:fig10}(a)).
The accumulation of ${\langle{\Gamma}\rangle}_{\Omega}$ induces a deviation of the shear-layer, confirmed by the integrated cross-stream velocity ${\langle{v}\rangle}_{\Omega}$ inside an extended domain $\Omega$ limited by $(x/H,y/H)=([0,0.07],[1.15,1.19])$ displayed in figure~\ref{fig:fig10}(b).
When the actuation is started, the downward cross-stream velocity significantly raises to values of about $8\,\%$ of $U_o$.	
The fluctuations of both ${\langle{\Gamma}\rangle}_{\Omega}$ and ${\langle{v}\rangle}_{\Omega}$ indicate that the shear-layer flaps due to the high-frequency injection of circulation leading to the periodicity of $C_{p_T}$.

\begin{figure}
				        \centering
                \includegraphics[scale=0.8]{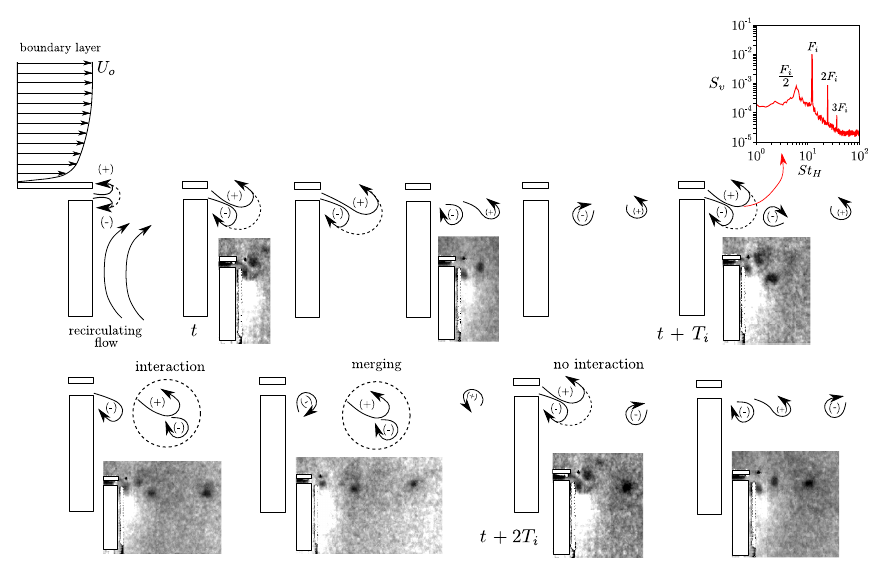}
                \caption{Cyclic evolution of the pulsed jet formation at high-frequency forcing. The picture shows the vortex formation and its interactions along two actuation cycles. The power spectral density $S_v$ of the cross-stream velocity at $(x/H,y/H)=(0.03,1.16)$ contains peaks at half the forcing frequency which indicates the existence of vortex merging also visible in the flow visualizations.}
                \label{fig:fig11}
         \end{figure}

Due to the limited resolution of our PIV system, we are not able to identify the core of the pulsed vortex.
Its length scale is of the same order as the actuation slit width ($\sim 1\text{mm}$).
Yet, flow visualizations acquired using this set-up shed some light on the phenomenology of this forcing illustrated in figure~\ref{fig:fig11}.
			 
Two counter-rotating vortices emerge from the exit slit at the beginning of the actuation cycle.
The corresponding jet structures closely resemble the Schlieren visualizations of synthetic jets during the blowing phase obtained by \citet{Smith05}.
Their averaged diameter scales with the slit width $h\sim1\text{mm}$.
Since the convection velocity adjacent to each of these structures is different (i.e. boundary layer at one side and recirculating flow just below the jet), the lower clockwise rotating vortex stays close to the slit outlet while the upper structure is convected with $V_c\sim 0.5U_o$.
Interestingly, at the beginning of the forthcoming actuation period ($t+T_i$), the new pulsed jet interacts with the preceding clockwise vorticity, resulting in a merged structure which is further convected downstream.

Power spectral density ($S_v$) of the cross-stream velocity at $(x/H,y/H)=(0.03,1.16)$ reveals vortex pairing by the presence of the first actuation subharmonic $F_i/2$ (see insert figure).
Similarly, the next clockwise vortex still remains first close to the edge and is convected downstream towards the end of the actuation period.
In the following cycle ($t+2T_i$), however, there is no interaction between the structures.
The entire process leads to a periodic accumulation of circulation close to the edges, promoting shear-layer deviation towards the center of the separated flow.
		        
\begin{figure}
				        \centering
                \includegraphics[scale=0.42]{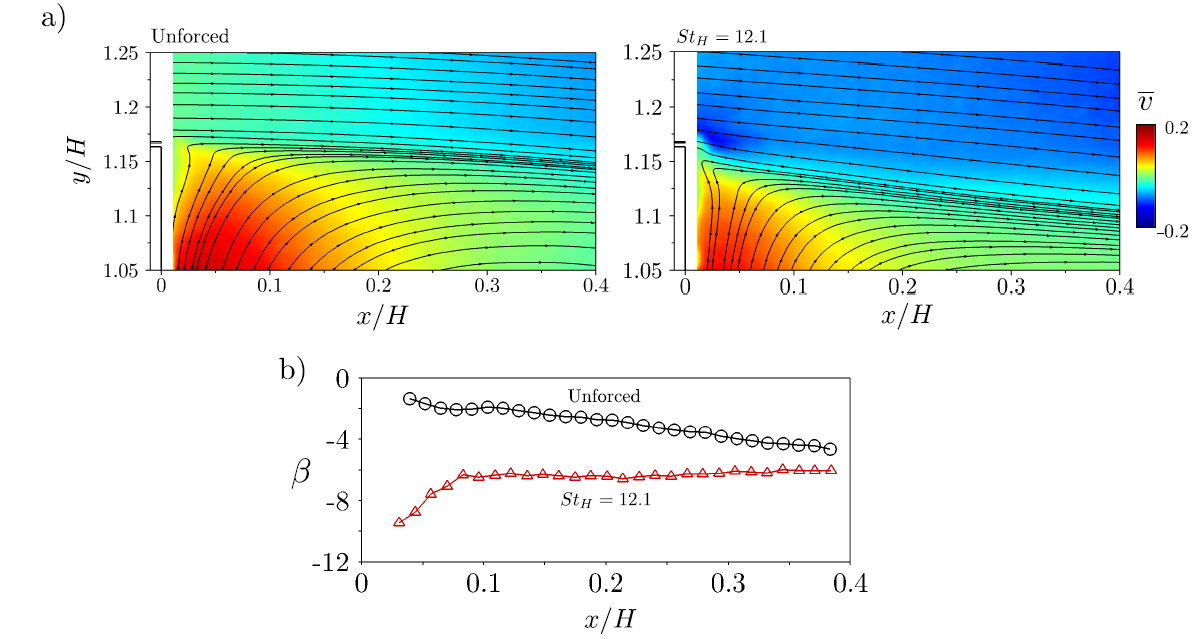}
                \caption{Streamlines and distribution of time-averaged cross-stream velocity $\overline{v}$ at ${Re_{H}}=3 \times 10^5$. The actuation parameters are $St_H=12.1$ and $C_\mu=4.6 \times 10^{-3}$ ($L_o\sim12.8$ and $V_{j_{\text{eff.}}}\sim9.1\,\text{m}\text{s}^{\text{-}1}$). c) Velocity angle $\beta$ along the separating streamlines for both unforced and forced wakes.}
		    \label{fig:Comparison_Mean_V_Nat_HF_650mbar}
        \end{figure}

The resulting time-averaged shear-layer is illustrated in figure~\ref{fig:Comparison_Mean_V_Nat_HF_650mbar}(a) by the streamlines together with the mean cross flow velocity $\overline{v}$.
An overall increase of the downward velocity magnitude is clearly noted by the flow streamlines.
Interestingly, the region of upward flow adjacent to the body surface (indicated by red color in the maps) is significantly reduced, suggesting reduction of local entrainment along the shear-layers and large-scale engulfment at the end of the bubble region.

In figure~\ref{fig:Comparison_Mean_V_Nat_HF_650mbar}(b), we compute the velocity angle $\beta$ along the separating streamlines originated at $(x/H,y/H)=(0,1.17)$.
The reference flow presents low velocity deviation close to the edge and the streamlines are only slightly directed downwards.
The minimum velocity angles of -$4^\circ$ are calculated at $x/H\sim0.3$.
On the other hand, the forcing strongly deviates the flow at the very beginning of the shear-layer with angles of -$10^\circ$ gradually decreasing along $x$.
We can deduce from this streamwise evolution that significant changes of the velocity field curvature take place, modifying the pressure gradients across the shear-layer.
As an analogy to the passive control discussed in $\S\,\,1$, this averaged scenario points to a \textit{fluidic boat-tailing} effect where the pulsating jets deviate flow without geometric modifications, resulting in base pressure recovery and drag reduction.
It is worth to mention here that flow deviation has been also noted along the lower and lateral shear-layers \citep{Barros14}, and we focused on this effect close to one edge of the model with the high-speed PIV apparatus. 

\subsection{Influence of the jet amplitude} 		   

\begin{figure}
				        \centering
                \includegraphics[scale=0.38]{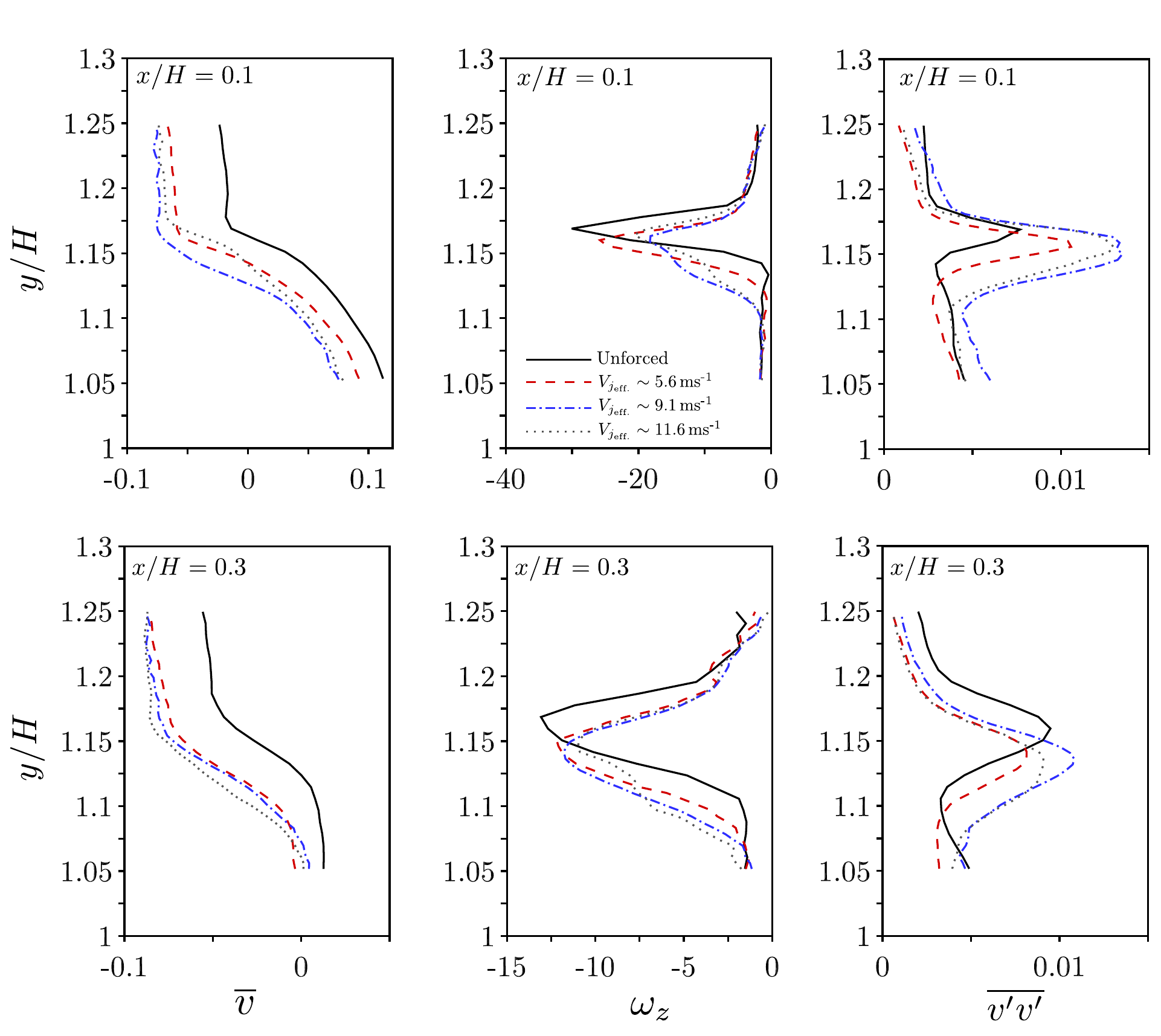}
                \caption{Influence of jet amplitude for high-frequency actuation ($St_H=12.1$) at ${Re_{H}}=3 \times 10^5$. First row: (from left to right): profiles at $x/H=0.1$ of the time-averaged cross-stream velocity $\overline{v}$, time-averaged vorticity $\omega_z$ and cross-stream velocity fluctuations $\overline{v'v'}$ when $V_{j_{\text{eff.}}}\sim\{5.6,9.1,11.6\}\,\text{m}\text{s}^{\text{-}1}$ corresponding to stroke lengths $L_o=\{8.2,12.8,15.1\}$. Second row: similar effects for a downstream streamwise position of $x/H=0.3$.}
                \label{fig:fig13}
        \end{figure}

The flow deviation depends on the introduced circulation close to the edge and is related to the jet strength.
We compare the effects of three forcing amplitudes $C_\mu=1.7 \times 10^{-3}$ ($V_{j_{\text{eff.}}}\sim5.6\,\text{m}\text{s}^{\text{-}1}$), $C_\mu=4.6 \times 10^{-3}$ ($V_{j_{\text{eff.}}}\sim9.1\,\text{m}\text{s}^{\text{-}1}$) and $C_\mu=7.5 \times 10^{-3}$ ($V_{j_{\text{eff.}}}\sim11.6\,\text{m}\text{s}^{\text{-}1}$) at fixed actuation frequency ($St_H=12.1$) and Reynolds number (${Re_{H}}=3 \times 10^5$).
Profiles of the $\overline{v}$ velocity, the vorticity $\omega_z$ and the cross-stream fluctuations $\overline{v'v'}$ are compared in figure~\ref{fig:fig13} at two streamwise positions along the upper shear-layer ($x/H=\{0.1,\,0.3\}$). 

Given the decrease of $\overline{v}$ at $x/H=0.1$, a deflection of the shear-layer is observed for all cases, but it is reduced for the smaller jet amplitude.
On the other hand, an increase of $V_{j_{\text{eff.}}}$ to $11.6\,\text{m}\text{s}^{\text{-}1}$ promotes smaller deviation, in agreement to the optimal value of $V_{j_{\text{eff.}}}\sim9.1\,\text{m}\text{s}^{\text{-}1}$ discussed in $\S\,\,4.2$.
Further downstream at $x/H=0.3$, the profiles look similar for these jet amplitudes.

Additionally to the flow deviation, a significant decrease of the minimum time-averaged spanwise vorticity $\omega_z$ is measured at $x/H=0.1$.
This is accompanied by an enlargement of the vorticity thickness, being higher for the optimal jet velocity. 
Both effects may decrease the spatial growth rate of the shear-layer modes in the linear stability framework \citep{Ho84}, as will be further illustrated by the streamwise evolution of the vorticity (see figure~\ref{fig:fig15}).
Finally, the increase of $\overline{v'v'}$ at $x/H=0.1$ is explained by the pulsed jet dynamics close to the slit edge.
An interesting feature, however, appears downstream ($x/H=0.3$), where a recovery and even diminution of $\overline{v'v'}$ along the shear layer is noticed.
Hence, the enhanced entrainment due to the formation process of the pulsed jet is followed by a relaxation of velocity fluctuations further downstream.

In figure~\ref{fig:fig13}, it is worth to mention that these three forced wakes differ only $3\,\%$ in base pressure at ${Re_{H}}=3 \times 10^5$.
Hence, the modifications of these quantities are not as significant as their changes when compared to the natural flow.
Meanwhile, we markedly note a lower vectoring effect and a smaller decrease of peak vorticity for the lowest jet amplitude, both of them responsible for their base pressure variation.

Part of the ${\langle{\Gamma}\rangle}_{\Omega}$ circulation previously analyzed comes from the pulsed jet circulation $\Gamma_j$.
The circulation flux issued from the exit slit may vary as $(\Gamma_j)_{\text{flux}}\propto{(V_{j_{\text{eff.}}})}^2$, further emphasizing that higher jet velocities are capable to induces more deviated flow.
However, the existence of an optimum jet amplitude on the $\gamma_p$ evolution contradicts this tendency.
The studies of \citet{Oxlade15} show that this optimality is correlated to a smaller peak vorticity of the jet structures shown in figure~\ref{fig:fig11}.
It is natural to conjecture that the dynamics of the jet structures might have an additional influence in this process.

\subsection{Reynolds number effects and shear-layer growth} 		   
		
In this section, we investigate how actuated flow is affected by varying the upstream Reynolds number ${Re_{H}}=\{2,3,4\} \times 10^5$: we maintain fixed both the forcing frequency ($F_i=610\,\text{Hz}$) and the pulsed jet amplitude ($L_o=8.2$ and $V_{j_{\text{eff.}}}\sim5.6\,\text{m}\text{s}^{\text{-}1}$).
This configuration corresponds to a 11\% drag reduction at ${Re_{H}}=3 \times 10^5$.
		
\begin{figure}
				        \centering
                \includegraphics[scale=0.44]{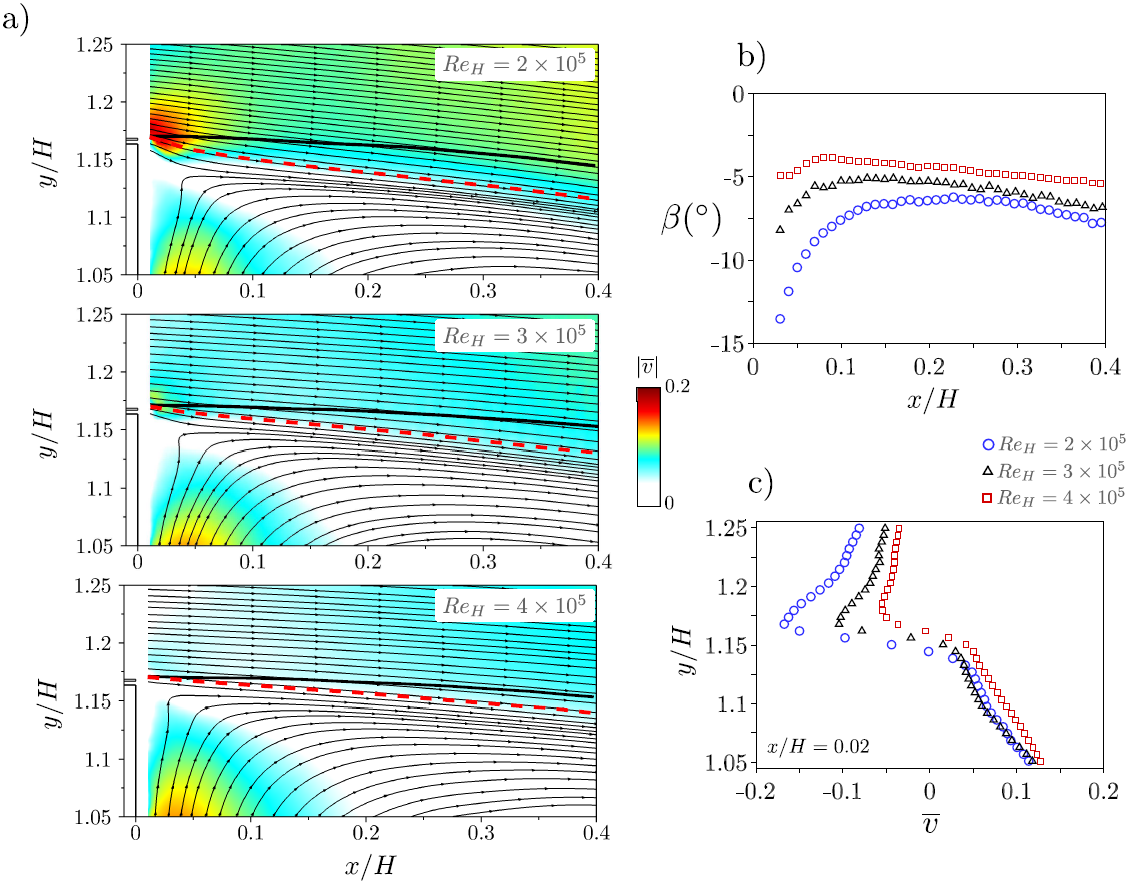}
                \caption{Effects of upstream Reynolds number on the forced shear-layer deviation with $L_o\sim8.2$ ($V_{j_{\text{eff.}}}\sim5.6\,\text{m}\text{s}^{\text{-}1}$) and high-frequency actuation. a) Cross-stream velocity distribution $|\overline{v}|$ and streamlines for varying $Re_H$. b) Streamwise evolution of the velocity angle $\beta$ along the separating streamlines indicated in a). c) Cross-stream velocity profile at $x/H=0.02$ presenting higher flow deviation at lower Reynolds number.}
                \label{fig:fig14}
        \end{figure}

As shown in figure~\ref{fig:fig14}(a), the distribution of $|\overline{v}|$ and the flow streamlines are highly influenced by $Re_H$.
The comparison between the unforced (full lines) and forced (dashed lines) streamlines reveals that flow deviation becomes more significant for lower $U_o$. 
The streamwise evolution of the deviation angle $\beta$ shown in figure~\ref{fig:fig14}(b) indicates curvature changes of the mean streamlines, which might explain the higher base pressure recovery at lower upstream velocities.
In figure~\ref{fig:fig14}(c), the increased downward velocity in the initial part of the shear-layer is attested by velocity profiles ($x/H=0.02$).
In the region $y/H\!>\!1.17$, the entire boundary layer flow is deviated and $\overline{v}$ is more than doubled when $Re_H$ is reduced to one-half.
Higher flow deviations at low free-stream velocities were also found in the experiments of \citet{Smith02}.
However, in this former case a main jet was forced by adjacent synthetic jets, in contrast to the present results for a wake flow.

\begin{figure}
				        \centering
                \includegraphics[scale=0.35]{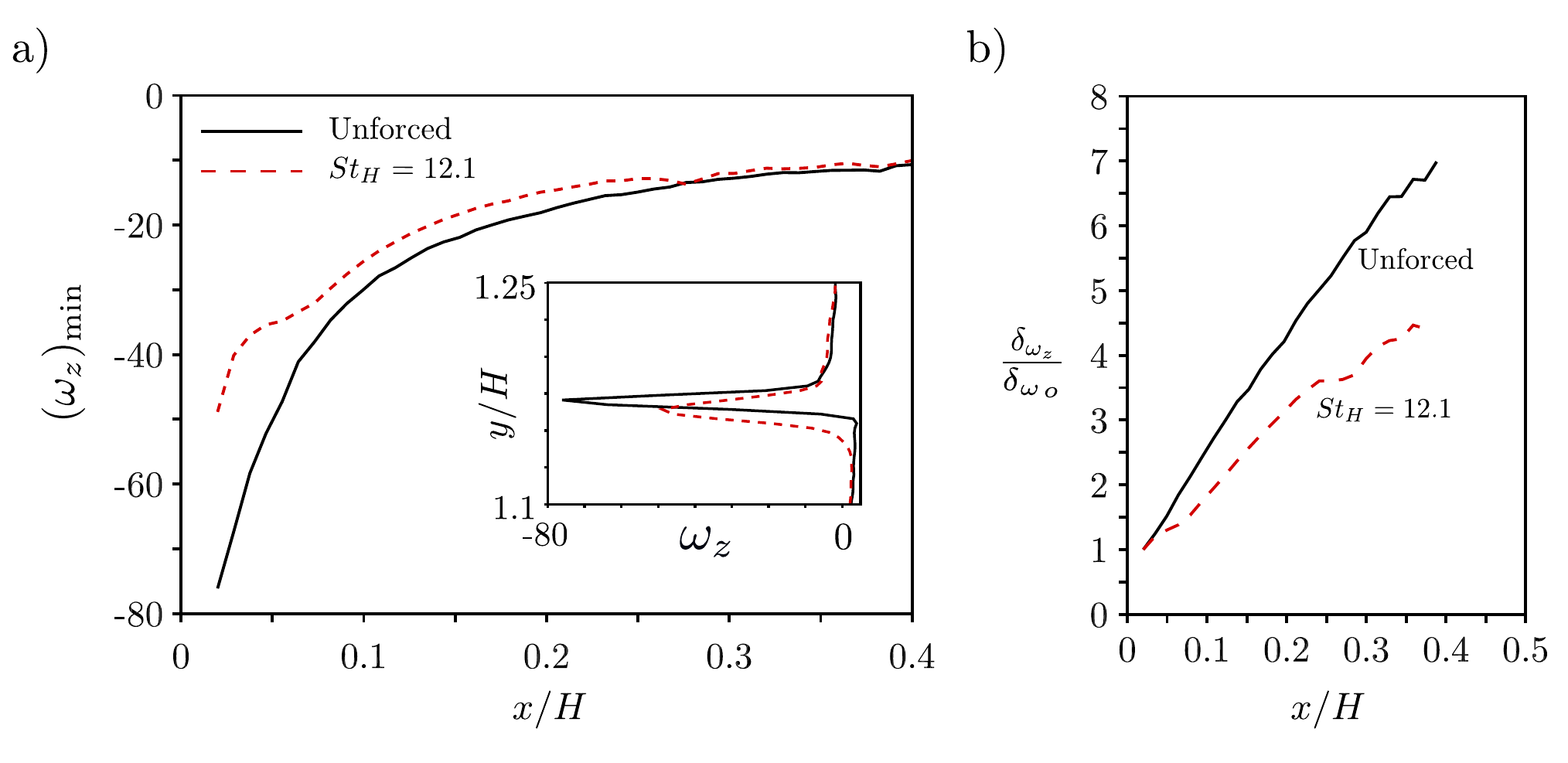}
                \caption{Decay of spanwise vorticity along the streamwise direction at ${Re_{H}}=3 \times 10^5$. a) Decay of the peak vorticity $(\omega_z)_{\text{min}}$ and vorticity profile at $x/H=0.02$ close to the model edge for unforced and actuated flows. The forcing parameters are $St_H=12.1$ and $L_o\sim8.2$ ($V_{j_{\text{eff.}}}\sim5.6\,\text{m}\text{s}^{\text{-}1}$). b) Streamwise growth of the vorticity thickness $\delta_{\omega_z}$, where ${\delta_{\omega_o}}$ is computed for the unforced and forced configurations separately at $x/H=0.02$.}
                \label{fig:fig15}
        \end{figure}

The flow deviation is accompanied by a change of vorticity spread and growth along the shear-layer.
Figure~\ref{fig:fig15}(a) shows the streamwise evolution of the peak vorticity $(\omega_z)_{\text{min}}$.
The unforced mean flow vorticity decays as ${\omega_z}\propto{x^{-0.72}}$, in agreement with turbulent separated boundary layer measurements from \citet{Morris03} exhibiting an exponent of -0.79.
When control is applied, there is an overall decrease of the peak vorticity confirmed by a profile at $x/H=0.02$: $(\omega_z)_{\text{min}}$ reduces by more than 30\% at this location, but further downstream ($x/H>0.25$) both evolutions collapse.
		
A diminution of $(\omega_z)_{\text{min}}$ along $x$ is associated to an enlargement of the vorticity thickness $\delta_{\omega_z}$, defined as:

\begin{equation}
{\delta_{\omega_z}}=\frac{U_{\text{max}}-U_{\text{min}}}{{|\frac{\partial{\overline{u}}}{\partial{y}}|}_{\text{max}}}.
\end{equation}		
				
In figure~\ref{fig:fig15}(b), the streamwise evolution of $\delta_{\omega_z}/{\delta_{\omega_o}}$ clearly shows a decrease of the shear-layer growth rate, despite the initial increase by forcing of ${\delta_{\omega_o}}$ computed very close to the trailing edge at $x/H=0.02$.
The robustness of these results has been attested for all Reynolds numbers \citep{Barros15a}.

\begin{figure}
				        \centering
                \includegraphics[scale=0.43]{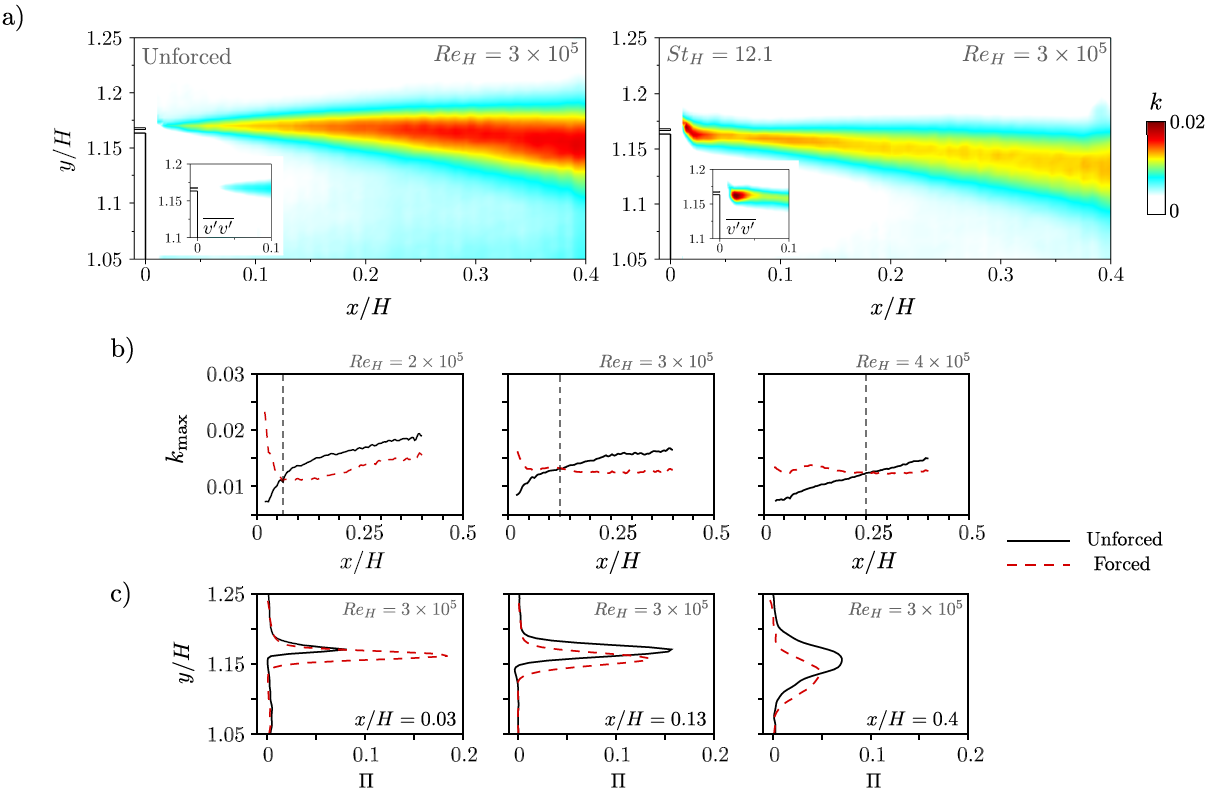}
                \caption{Turbulent kinetic energy $k$ and production $\Pi$ along the upper shear-layer. a) Contour maps of $k$ at ${Re_{H}}=3 \times 10^5$. The inserted picture indicates the cross-stream velocity fluctuations $\overline{v'v'}$ near the exit slit. b) Streamwise evolution of $k_{\text{max}}$ for varying $Re_H$. The vertical dashed lines are located at $x=0.07,0.13,0.25$ corresponding to the positions where both unforced and actuated flow present the same $k_{\text{max}}$. c) Cross-stream profiles of turbulent kinetic energy production $\Pi$ at three streamwise locations for ${Re_{H}}=3 \times 10^5$. The jet parameters are $L_o\sim8.2$ and $V_{j_{\text{eff.}}}\sim5.6\,\text{m}\text{s}^{\text{-}1}$.}
                \label{fig:fig16}
        \end{figure}

This reduction of the growth rate is linked to a damping of the turbulent kinetic energy $k=0.5\times({\overline{u'u'}}+{\overline{v'v'}})$ - more precisely, the two contributions to $k$ measured with the PIV system - displayed in figure~\ref{fig:fig16}(a).
The unforced, natural evolution of the shear-layer is associated to a streamwise raise of $k$. 
Forcing produces an increase of turbulent kinetic energy up to $x/H\sim0.13$, while more downstream $k$ is damped.
The initial growth of $\overline{v'v'}$ is particularly visible in the vicinity of the edge close to the pulsed jet emission. 

Figure~\ref{fig:fig16}(b) illustrates the increase of $k_{\text{max}}$ promoted by actuation and successive damping downstream a critical position, where natural and forced configurations present equivalent values.
This location is displaced downstream from $x/H=0.07$ at the lowest Reynolds number up to $x/H=0.25$ at the highest upstream velocity.
Further downstream, $k_{\text{max}}$ is smaller compared to the unforced flow counterpart, indicating a stabilization of the cross-stream dynamics.
In the time-averaged framework, the evolution of $k$ is dictated by the production of turbulent kinetic energy $\Pi(x,y)$, whose in-plane contribution is:

\begin{equation}
{\Pi}=-{\overline{u'u'}{\left(\frac{\partial{\overline{u}}}{\partial{x}}\right)}}-{\overline{v'v'}{\left(\frac{\partial{\overline{v}}}{\partial{y}}\right)}}-{\overline{u'v'}{\left(\frac{\partial{\overline{u}}}{\partial{y}}+\frac{\partial{\overline{v}}}{\partial{x}}\right)}}.
\label{subeq:4}
\end{equation}

Profiles of $\Pi$ at ${Re_{H}}=3 \times 10^5$ are shown in figure~\ref{fig:fig16}(c).
The selected streamwise positions correspond respectively to the beginning of the shear-layer, where the actuation significantly increases $k$; the position at which $k_{\text{max}}$ is approximately the same for both unactuated and actuated flows; and finally a location where the damping of the velocity fluctuations is clearly visible at the end of the PIV domain.
Close to the jet origin, the production term of the actuated flow has more than twice as high peak values of the reference case.
Then, $\Pi$ is reduced further downstream.
The highest reduction of $\Pi$ of about 30\% is observed for ${Re_{H}}=\{2,3\} \times 10^5$ while for ${Re_{H}}=4 \times 10^5$ it is reduced to 15\%.

Similar stabilizing behavior of shear-layers under high-frequency forcing has been previously reported.
With the help of linear stability analysis of the time-averaged streamwise velocity profiles, \citet{Dandois07} found that high-frequency forcing produces a drop of the growth rates for the most amplified shear-layer modes.
This is confirmed in the present study by the increased shear-layer thickness.
On the other hand, \citet{Vukasinovic10} showed the presence of high-dissipation zones in the vicinity of the actuator and therefore claimed that this zone drains energy from the mean flow leading to a stabilization, suggesting a dissipative small scale actuation.
Our results seem to confirm both analysis for a bluff body wake and indicate that such mechanisms, coupled to the fluidic boat-tailing effect, alter the global balance of the recirculating flow leading to pressure drag reduction.

\subsection{Global wake modifications} 		   

\begin{figure}
				        \centering
                \includegraphics[scale=0.43]{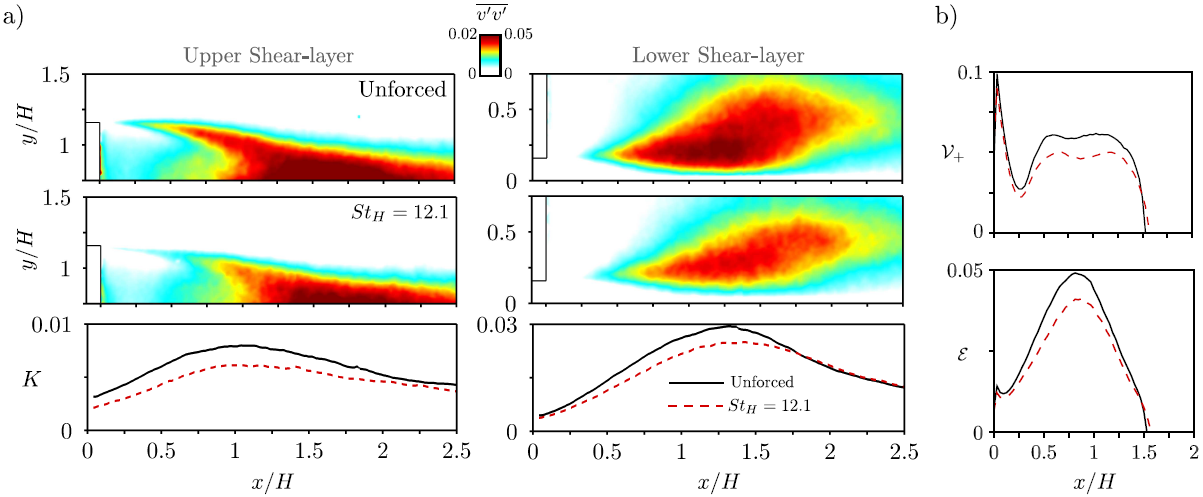}
                \caption{Velocity fluctuations in the near wake and streamwise evolution of the entrainment measure. a) Contour maps of $\overline{v'v'}$ for both the upper and the bottom shear-layers. Streamwise evolution of $K$. b) Streamwise evolution of ${\mathcal{V}_{+}}$ and ${\mathcal{E}}$ inside the region of reverse flow up to the end of the recirculating region limited by $x=L_r\sim1.5$. The forcing conditions are $St_H=12.1$ and $L_o\sim8.2$ ($V_{j_{\text{eff.}}}\sim5.6\,\text{m}\text{s}^{\text{-}1}$) for an upstream Reynolds number ${Re_{H}}=3\times 10^5$.}
                \label{fig:fig17}
        \end{figure}

Up to now, we focused on the effects of high-frequency forcing on shear-layer deviation and its growth rate close to the edges of the model.
A global change of the whole wake dynamics occurs due to the reduction of the shear growth, and an extension of our analysis to the entire recirculating flow appears necessary.
We present in figure~\ref{fig:fig17}(a) the cross-stream velocity fluctuations $\overline{v'v'}$ and the turbulent kinetic energy integrated vertically across any selected region of the flow:

\begin{equation}
{K(x)}=\int_{y_{\text{min}}}^{y_{\text{max}}} {k}(x,y)dy.
\end{equation}

Both quantities are shown in two regions around the top and the bottom shear-layers.
Not only the reduction of $K$ is observed along the upper part, as reported before using time-resolved PIV, but also along the bottom development of the wake.
Similar trends are obtained whatever the Reynolds number, but a more pronounced decrease of velocity fluctuations is confirmed for the lowest upstream velocity corresponding to ${Re_{H}}=2 \times 10^5$. 

The reduction of mixing along the shear-layers should lead to lower flow entrainment inside the recirculating zone.
A physical measure of the entrained reverse flow represents the integral of $|\overline{v}|$ and the total kinetic energy along $y$ inside the domain $\Omega_{\overline{u}<0}$:
 				
\begin{equation}
{\mathcal{V}_{+}(x)}=\int_{\Omega_{\{\overline{u}<0\}}} {|\overline{v}}(x,y)|dy,
\end{equation}

\begin{equation}
{\mathcal{E}(x)}=\int_{\Omega_{\{\overline{u}<0\}}}{\left({\frac{{\overline{u}}^2+{\overline{v}}^2+{\overline{u'u'}}+{\overline{v'v'}}}{2}}\right)}dy.
\end{equation}

These quantities are displayed in figure~\ref{fig:fig17}(b) at Reynolds number ${Re_{H}}=3 \times 10^5$.  
Actuation leads to a decrease of the cross-stream motion and the total kinetic energy of the reverse flow.
Similar behavior is measured at other Reynolds numbers.
Close to the model, a decrease of ${\mathcal{V}_{+}}$ appears highlighting a diminution of the upward flow adjacent to the rear surface.
This points to a decrease of the entire recirculating flow intensity, and, thus of the whole kinetic energy inside the bubble.
It is important to remark that the analyzed quantities are computed in the symmetry plane of the wake which can only give an order of magnitude of the really entrained flow, since in reality it is three-dimensional.

\begin{figure}
				        \centering
                \includegraphics[scale=0.18]{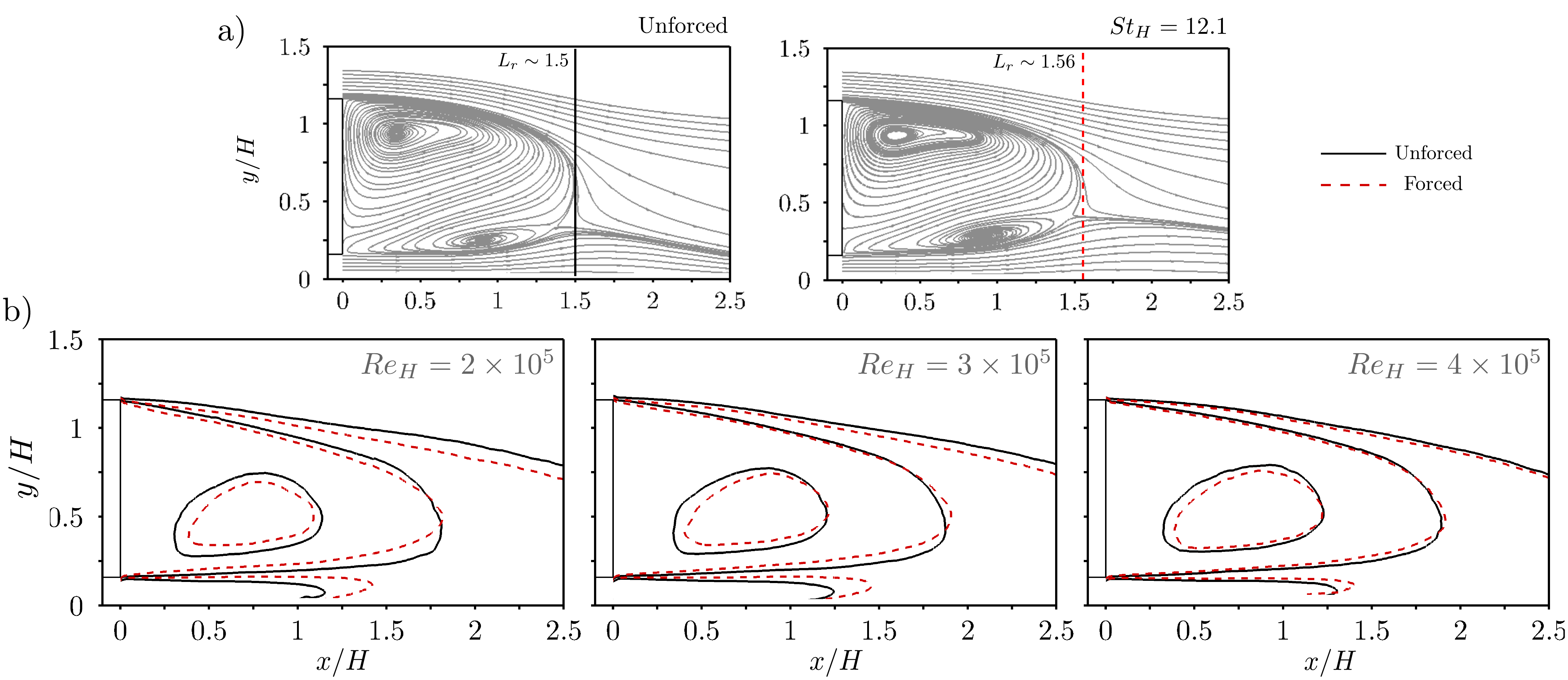}
                \caption{Global wake topology and effects of high-frequency actuation. a) Streamlines of the in-plane velocity field (${Re_{H}}=3 \times 10^5$). b) Streamwise velocity contours $\overline{u}=\{-0.25, 0.25, 0.65\}$ indicating a narrowing of the wake at different Reynolds numbers. The actuation amplitude corresponds to $L_o\sim8.2$ ($V_{j_{\text{eff.}}}\sim5.6\,\text{m}\text{s}^{\text{-}1}$). }
                \label{fig:fig18}
        \end{figure}

We turn our attention now to the modifications of the wake topology.				
The streamlines of the natural and forced wakes at ${Re_{H}}=3 \times 10^5$ are depicted in figure~\ref{fig:fig18}(a).
The limit of the bubbles corresponds to the recirculation length $L_r$.
Despite of a base pressure increase of $15\,\%$, only a very slight increase of the mean recirculation bubble length in the symmetry plane is measured for actuation at $St_H=12.1$ with $L_o\sim8.2$ ($V_{j_{\text{eff.}}}\sim5.6\,\text{m}\text{s}^{\text{-}1}$).
The recirculating bubble length is increased by at most $2-4\%$ considering all upstream velocities.
In comparison with the low-frequency actuation analyzed in the Appendix A, the studied forcing at $St_H\sim0.8$ leads to a decrease of $12\,\%$ on $\langle{\overline{C_p}}\rangle$ and $18\,\%$ on the bubble length.
Surprisingly, we do not observe significant changes of the mean streamlines of the wake despite a drag decrease of $11\%$.

A careful analysis of the streamwise velocity $\overline{u}$, however, sheds light on the subtile changes of the wake geometry.
In figure~\ref{fig:fig18}(b), the contour lines corresponding to iso-values of streamwise velocity $\overline{u}=\{-0.25,0.25,0.65\}$ are shown respectively for ${Re_{H}}=\{2,3,4\} \times 10^5$.
We observe a narrowing of the wake, which is more pronounced at ${Re_{H}}=2 \times 10^5$, in agreement with the highest flow deviations measured in the vicinity of the upper edge.
The flow deviation along the bottom shear-layer appears similar, suggesting an equal deviation by forcing along all slits, also
reported with respect to the lateral shear-layers \citep{Barros14}.

Finally, we study the forcing effects on the global wake mode measured downstream the end of the recirculating bubble.
From hot-wire measurements at $x/H=2$ and $x/H=3$ in the symmetry plane ($y/H=0.9$) and in a horizontal plane ($y/H=0.6$ and $z/H=0.5$) parallel to the ground, we concluded that no significant modifications are observed in the shedding frequencies, differently to what has been previously found at lower-actuation frequencies \citep{Barros16}.
However, a closer inspection reveals a slight decrease of the peak intensity in the spectrum with control.
It indicates that these modes are modestly damped and their detection is only possible in the rear part of the recirculating flow as observed by \citet{Grandemange13}.	

\subsection{A conceptual scenario for high-frequency drag reduction} 		   

The high-frequency forcing impacts the wake by two mechanisms: an initial flow vectoring narrows the whole wake and a reduction of the shear-layer growth rate diminishes the turbulent kinetic energy and entrainment.

\begin{figure}
				        \centering
                \includegraphics[scale=1.5]{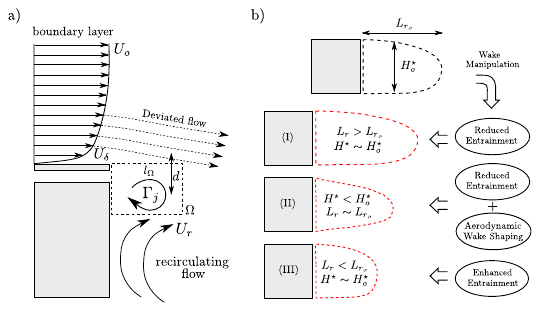}
                \caption{Conceptual sketch of the high-frequency actuation. a) Deviation effect on the shear-layer by injected circulation. b) Global effects by wake manipulation: reduced entrainment and bubble elongation (I), reduced entrainment coupled to bubble shaping for high-frequency forcing (II) and enhanced entrainment decreasing the bubble length for low-frequency forcing (III).}
                \label{fig:fig19}
        \end{figure}

The analysis of flow deviation dynamics produced by actuation highlights the importance of circulation flux in the vicinity of the rear edges, responsible for the deviation of the shear-layers.
Figure~\ref{fig:fig19}(a) shows the trailing edge and its vicinity filled with a patch of fluid containing clockwise circulation during actuation.
We assume the flow deviation depends on the enhanced entrainment up to a distance $l_\Omega$ from separation.
In this region $\Omega$, the injected circulating motion deviates the flow.

The circulation $\Gamma_{\Omega}$ induces the cross-stream velocity ${v_{\text{ind.}}}$ causing the flow deviation $\beta_{\text{ind.}}=({v_{\text{ind.}}}/{U_o})$.
By considering an analogy to the velocity induced by a potential vortex, we assume $\beta_{\text{ind.}}\propto{\Gamma_{\Omega}/({U_o}{d})}$, where $d$ is related to the slit centerline position with respect to the outer surface adjacent to the boundary layer.

In particular, $\Gamma_{\Omega}$ crucially depends on the organized circulation from each pulsed vortex $\Gamma_j$ and its convection outside $\Omega$.
Following \citet{Dabiri09}, we may write ${\Gamma_j}\propto{0.5{T_f}{{V_{j_{\text{eff.}}}}^2}}$, where $T_f$ is the vortex formation time.
The pulsed jet circulation will be convected with $U_c\propto \mathcal{F}({U_\delta},{U_r},{V_j})$ dependent on the boundary layer (${U_\delta}$) and the recirculating flow (${U_r}$) characteristic velocities, both increasing with $U_o$.
During a period of actuation $T_i$, the convection distance of vortices is about $l={U_c}{T_i}$. 
If $U_o$ increases or the actuation period raises, $l$ becomes larger than $l_\Omega$ and the average $\Gamma_{\Omega}$ decreases causing less flow deviation.

By coupling both the circulation and its convection, the dependence of ${\beta_{\text{ind.}}}$ can be expressed as:

\begin{equation}
{\beta_{\text{ind.}}}\sim \mathcal{F}\left[\left(\frac{{T_f}{{V_{j_{\text{eff.}}}}^2}}{{U_o}{d}}\right), \left(\frac{{l_\Omega}}{U_c T_i}\right)\right].
\label{subeq:11}
\end{equation}

Keeping all other quantities fixed, the increase of jet velocities up to the critical $V_{j_{\text{eff.}}}$ leads to an increase of $\Gamma_j$.
From our relation, this is directly linked to more deviation ${\beta_{\text{ind.}}}$ causing an amplified fluidic boat-tailing effect.
On the other hand, flow is convected more rapidly for higher $U_o$ and the pulsed jets leave $\Omega$, reducing the total circulation in this region and consequently ${\beta_{\text{ind.}}}$.
This is confirmed by the smaller vectoring angles $\beta$ measured at higher upstream flow velocities.

From the vortex formation studies with piston-cylinder arrangements, \citet{Gharib98} and \citet{Dabiri09} showed that ${T_f}\propto {h}{T^{\star}}/\overline{V_j}$, where $T^{\star}$ is universal for each pulsating geometry. 
This implies also that $h$ is implicitly contained in equation~\ref{subeq:11}.

An additional effect of increasing $V_j$ is the possibility to provoke vortex pinch-off, avoiding further transfer of circulation to $\Gamma_j$.
In this sense we could hypothesize the optimality of $L_o$ (or $V_{j_{\text{eff.}}}$) linked to vortex pinch-off as observed by \citet{Gharib98}.
Further measurements with systematic varying parameters should be performed to draw a more complete scenario.
The recent experimental results from \citet{Oxlade15} at fixed upstream flow conditions support our hypothesis by showing the correlation between the vorticity amount contained in the main vortex and the optimum drag and jet velocity.

Describing how actuation affects the entire bubble topology is necessary to understand base pressure and drag changes.
In general, a decrease of shear-layer growth is related to smaller wake entrainment rates, which elongates the recirculating flow as indicated by sketch (I) in figure~\ref{fig:fig19}(b). 
On the other hand, the initial shear-layer deviation alone would imply not only a narrowing of the wake but also a reduction of the recirculating length.
The balance of both effects might explain the very similar wake lengths $L_r\sim L_{r_o}$ for high-frequency actuation (II), where the reduced velocity fluctuations act to elongate the bubble length.
The forced bubble presents a higher aspect ratio ${L_r}/{H^{\star}}$ by the diminution of ${H^{\star}}$ from aerodynamic shaping, differently to the low actuation frequencies where an enhancement of the entrainment decreases ${L_r}$ shown by (III) (see Appendix A and \citet{Barros16}).

Both scenarios are consistent with Roshko's models on the wake equilibrium and base pressure \citep{Roshko55,Roshko93a,Roshko93b}.
In an analogy to the simplified two-dimensional wake, let us recall the streamwise mean momentum balance along the closing separating line of the bubble revisited by \citet{Balachandar97}:

\begin{equation}
\langle{\overline{C_{p_{b}}}}\rangle H=\int\limits_{\partial\mathcal{B}}{\overline{C_p}}({\vec{n}}\cdot{\vec{x}})ds + 2\int\limits_{\partial\mathcal{B}}{{\overline{u'u'}}}({\vec{n}}\cdot{\vec{x}})ds+ 2\int\limits_{\partial\mathcal{B}}{{\overline{u'v'}}}({\vec{n}}\cdot{\vec{y}})ds,
\label{eq:momentum}
\end{equation}        
where $\vec{x}$ and $\vec{y}$ are the normal vectors of the reference system, $\vec{n}$ the local normal vector to the differential $ds$ of the bubble boundary $\partial\mathcal{B}$.
Fluxes of mean momentum across the wake boundary are assumed negligible.

As discussed in \citet{Balachandar97}, the normal and shear contributions of the right-hand side of the equation are both significant for the wake equilibrium, their magnitude varying more or less depending on the bluff body characteristics.
When forcing is applied, a variation of base pressure would obey $\delta\langle{\overline{C_{p_{b}}}}\rangle\sim\delta(\overline{C_{p_{\partial \mathcal{B}}}},{\overline{u'u'}},{\overline{u'v'}})$.
However, by computing the Reynolds stress terms along $\partial\mathcal{B}$, we note that their variation nearly compensates mutually, as ${\overline{u'u'}}$ and ${\overline{u'v'}}$ induce, respectively, positive and negative contributions to $\langle\overline{C_{p_{b}}}\rangle$.
Then, at first order, one obtain $\delta\langle\overline{C_{p_{b}}}\rangle\sim\delta(\overline{C_{p_{\partial \mathcal{B}}}})$. 

The problem being reduced to determine $\delta(\overline{C_{p_{\partial \mathcal{B}}}})$, one would interpret the pressure along the boundary as a result of large pressure gradients across $\partial\mathcal{B}$, which depends on its local streamline curvature $R$ and the local variation of velocity fluctuations along $\vec{n}$:
 
\begin{equation}
\frac{\partial{\overline{p}}}{\partial{n}}=\rho\frac{\overline{u_s}^2}{R}-\rho\frac{\partial{\overline{{v_n'}{v_n'}}}}{\partial{n}},
\label{eq:pressuregrad}
\end{equation} 
where $s$ is the coordinate tangential to the streamline and terms negligible compared to ${\overline{u_s}^2}$ are neglected \citep{Bradshaw73}.
It is clear from this formalism that changes of the wake aspect ratio would imply favorable pressure gradients by modifying $R$ along $\partial\mathcal{B}$.
Concomitantly, the damping of velocity fluctuations across $\vec{n}$ would similarly decreases $\frac{\partial{\overline{p}}}{\partial{n}}$ leading to a higher local bubble pressure $\delta(\overline{C_{p_{\partial \mathcal{B}}}})$.

\section{Additional drag reduction by unsteady Coanda blowing} 		   
	
To quantify the role of wake shaping on the pressure gradient along the bubble, we enhance flow deviation by adding small passive appendices to take advantage of the Coanda effect.
As discussed in $\S\,\,2$, a quarter cylinder with radius $r=9h$ is installed flush to the outlet of the jet slits.
The choice of this geometry is based on previous experiments in which the Coanda effect has been implemented for drag reduction in axisymmetric \citep{Freund94, Abramson11} and square-back \citep{Englar01,Pfeiffer12} geometries.
Our aim here is to characterize the unsteady Coanda forcing and the additional drag gain obtained by further flow deviation. 

\subsection{Impact on the base pressure and wake flow}

\begin{figure}
				        \centering
                \includegraphics[scale=0.35]{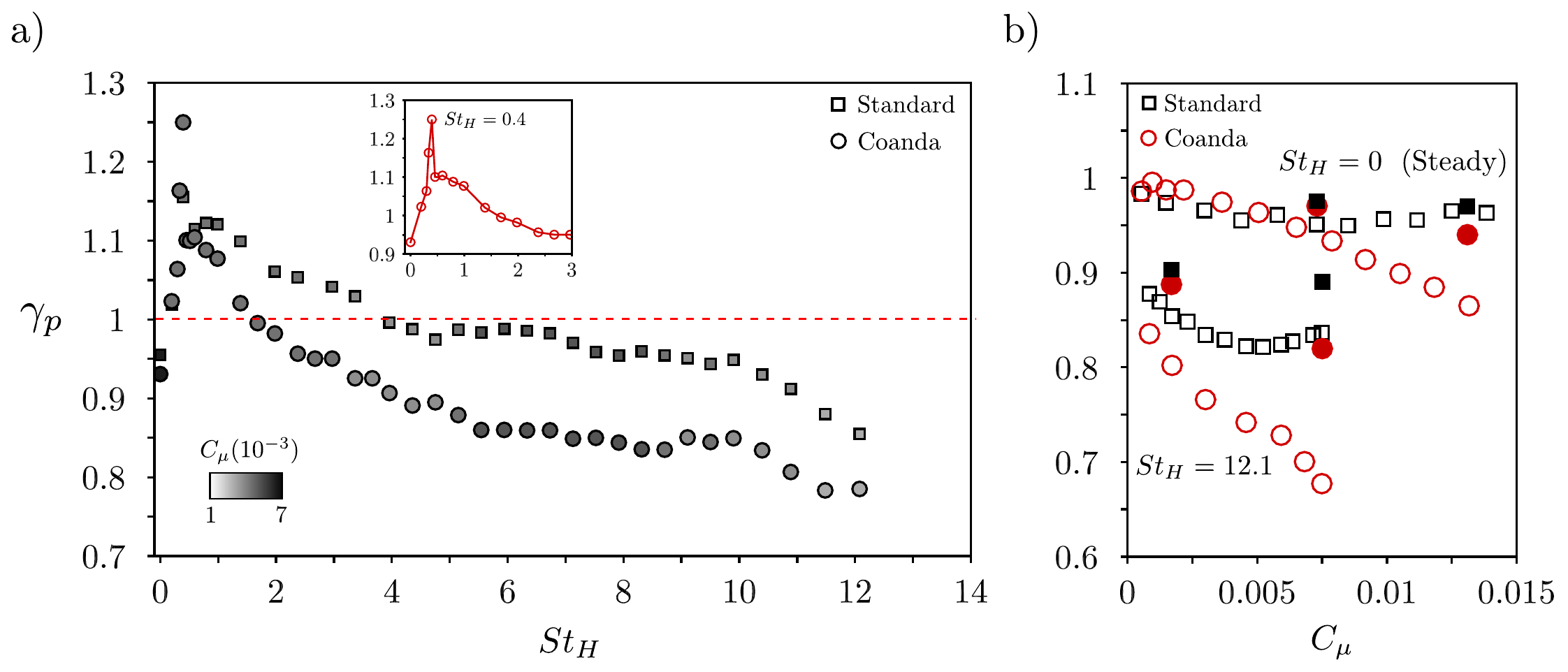}
                \caption{Impact of unsteady Coanda blowing on the base pressure parameter $\gamma_p$. a) Comparison between the standard actuation and the Coanda blowing by varying the actuation frequency $St_H$ at ${Re_{H}}=3.0\times 10^5$. The measurements are taken with a fixed input pressure $P_i$ of 1.45 bar. b) Effects of the momentum coefficient $C_\mu$ for the steady ($St_H=0$) and high-frequency forcing ($St_H=12.1)$ on $\gamma_p$ (open symbols) and $\gamma_d$ (filled symbols) selected for some configurations.}
                \label{fig:fig20}
        \end{figure}

In figure~\ref{fig:fig20}(a), the effects of unsteady Coanda forcing are characterized by the evolution of $\gamma_p$ as a function of $St_H$ at constant ${Re_{H}}=3.0\times 10^5$.
The Coanda blowing obeys a similar trend when compared to the standard case of straight pulsed jets.
The discrete peak with minimal base pressure at $St_H\sim0.4$ is preserved as shown in the insert.
Although the range of frequencies decreasing the base pressure remains comparable to the standard actuation, it is reduced to the interval $St_H\in[0.1,1.5]$.
Interestingly, base pressure recovery starts earlier from $St_H\sim2.0$ and extends to $St_H\sim12$: for the tested jet amplitudes, the Coanda forcing outperforms the straight jets lowering the values of $\gamma_p$. 

To quantify the influence of the jet amplitude on $\gamma_p$, we compare with and without the Coanda effect the steady ($St_H=0$) and the high-frequency ($St_H=12.1$) blowing configurations with variable $C_\mu$.
In addition, figure~\ref{fig:fig20}(b) indicates for some specific $C_\mu$ the measured drag and $\gamma_d$. 
Without curved appendices, steady blowing increases base pressure by approximately $5\,\%$, reducing the drag by at most $3\,\%$ in the range $C_{\mu}\in[0,0.015]$, corresponding to blowing velocities ${\overline{V_j}}/{U_o}\in[0,1.1]$.
This performance agrees with drag reductions obtained by previous studies with similar steady jet amplitudes in axisymmetric or square-back models \citep{Freund94,Wassen10,Krentel10}.
In the case of steady blowing, while no difference is measured up to $C_\mu\sim7.5\times{10^{-3}}$ (${\overline{V_j}}/{U_o}\sim0.55$), the Coanda effect improves base pressure recovery at higher blowing velocities.
For example, with $C_\mu=13.1\times{10^{-3}}$ corresponding to ${\overline{V_j}}/{U_o}\sim1.0$, base pressure increases by $14\,\%$ ($\gamma_p=0.86$) and drag decreases by $6\,\%$ ($\gamma_d=0.94$), similarly to what has been found by \citet{Pfeiffer12}.
The curve indicates that further increase of base pressure might be achieved using higher blowing velocities, a tendency in agreement with the measurements of \citet{Englar01} and \citet{Pfeiffer12}.

Using high-frequency Coanda blowing, we reached an even higher base pressure increase of 33\% ($\gamma_p=0.67$) at the largest $C_\mu=7.5\times{10^{-3}}$, and a total drag reduction of $18\,\%$.
Curiously, no saturation of $\gamma_p$ in the high-frequency Coanda actuation exists: one may therefore hypothesize that the shear-layer deviation is mainly dictated by the new imposed boundary condition, modifying the relation to the pulsed jet circulation discussed previously.			
				
\begin{figure}
				        \centering
                \includegraphics[scale=0.36]{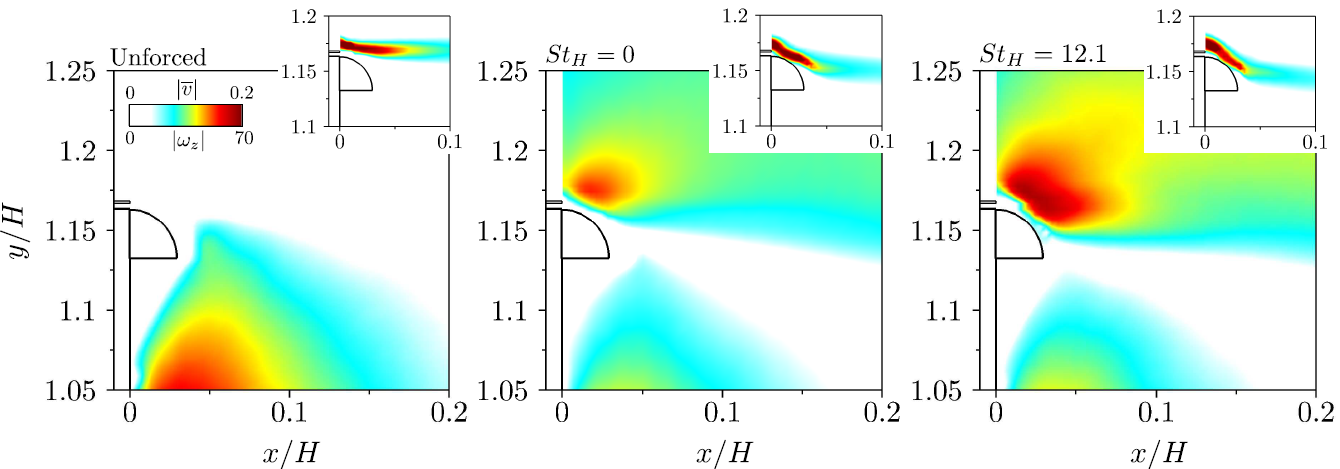}
                \caption{Impact of Coanda effect on the time-averaged cross-stream velocity $|\overline{v}|$ and vorticity $|\omega_z|$ fields. Comparison between the unforced flow (left), the steady forcing $St_H=0$ ($C_\mu=13\times{10^{-3}}$) and the high-frequency Coanda blowing $St_H=12.1$ ($C_\mu=7.5\times{10^{-3}}$).}
                \label{fig:fig21}
        \end{figure}

The flow deviation imposed by the Coanda effect is illustrated by the contour-maps of $\overline{v}$ in figure~\ref{fig:fig21}.
The unforced flow together with the steady blowing at $C_\mu=13\times{10^{-3}}$ ($\gamma_p\sim0.85$) and the high-frequency blowing at $C_\mu=7.5\times{10^{-3}}$ ($\gamma_p\sim0.67$) are compared.
The superiority of the unsteady forcing for flow deviation is clearly visible from both the vorticity $\omega_z$ and $\overline{v}$ distributions.
Moreover, we note a reduction of the upward velocity close to the rear surface of the model, similarly to the standard high-frequency configuration.

When comparing the steady blowing to the periodic actuation, the jet unsteadiness appears to be particularly important to flow attachment along the rounded surface.
Although the unsteady forcing with frequency $St_H=12.1$ presents a smaller averaged $C_\mu=7.5\times{10^{-3}}$, the instantaneous jet velocity may reach about three times its mean value, as shown in figure~\ref{fig:fig4}(a).
Hence, the time evolution of the momentum flux competing with the pressure gradient imposed by the rounded surface plays a significant role on the flow mechanisms leading to unsteady Coanda effect \citep{Jukes09}.
The use of steady blowing at velocities similar to the maximum jet amplitudes during the unsteady cycle would therefore be necessary in order to compare the effects of both actuation on flow attachment. Yet, this was not possible in the present study due to the needed input pressure $P_i$, beyond the limits of the pulsed jet system apparatus. Moreover, generating a steady jet for such control would require a large amount of power, considerably affecting the energy balance. This aspect is commented in what follows.

\begin{figure}
				        \centering
                \includegraphics[scale=0.42]{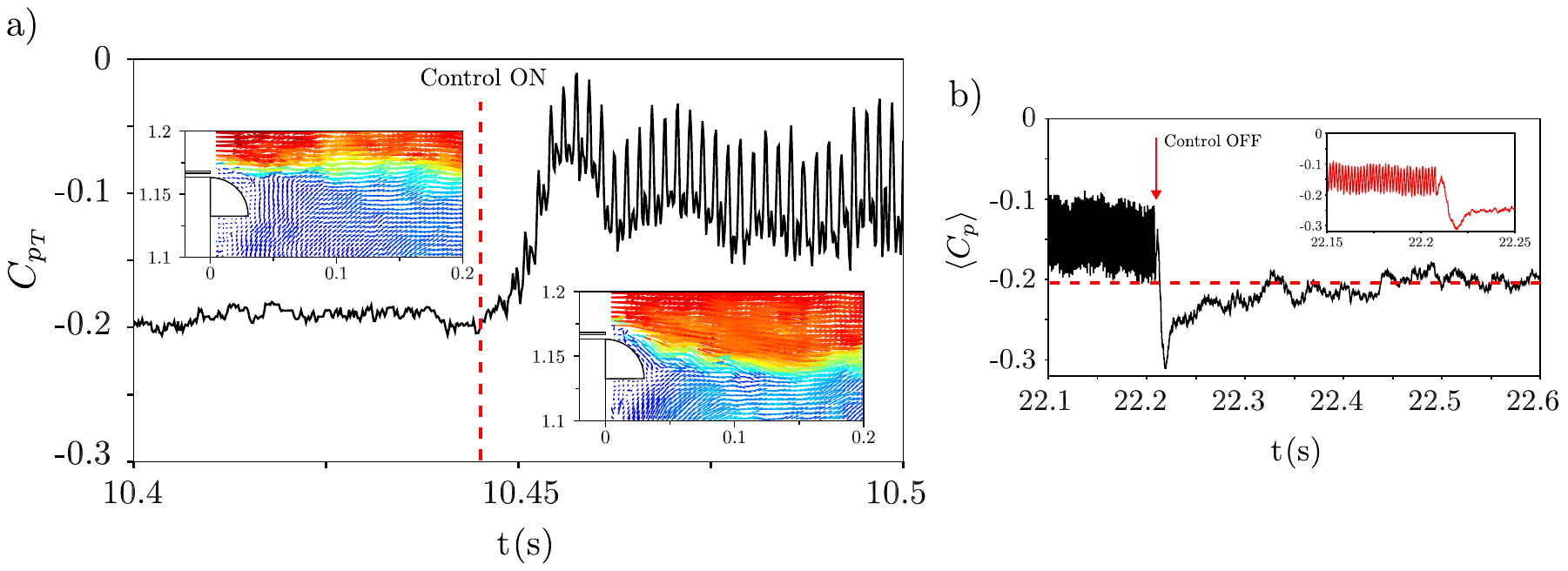}
                \caption{Transient analysis of high-frequency with Coanda effect at ${Re_{H}}=3.0\times 10^5$. a) Time-response of the top pressure sensor when control is applied at $St_H=12.1$ ($C_\mu=7.5\times{10^{-3}}$). b) Transient analysis when control is turned-off revealing the time-scales of the integrated base pressure $\langle{C_p}\rangle$ response. The horizontal dashed line represents the unforced pressure coefficient from longer time-averaged measurements.}
                \label{fig:fig22}
        \end{figure}

Let us describe the transient response of the high-frequency Coanda blowing by the time evolution of the top pressure coefficient $C_{p_T}$ depicted in figure~\ref{fig:fig22}(a) as well as the corresponding velocity vectors colored by the streamwise velocity $u$.
The total time window displayed is $0.1\,\text{s}$, which corresponds to 5 convective time units ($t^\star=t{U_o}/H$) or 1 shedding cycle from $f_{n_o}\sim0.2$.
The initial raise of $C_{p_T}$ takes approximately 6 actuation cycles ($\sim0.01\,\text{s}$) to be established, after which a decrease of pressure is noted due to the rapid stabilization of $P_i$ in the compressed air reservoir
\footnote{All the time-averaged measurements presented in this work are performed at least $5\,\text{s}$ following this stabilization, whose duration is roughly $0.05\,\text{s}$.}.
This time interval is equivalent to 0.5 wake convective units, indicating that the increase of pressure scales with the rapid wake adjustment by delaying flow separation from the Coanda effect.

The velocity snapshots emphasize the correlation between the increase of base pressure and an abrupt change of the velocity field curvature, as indicated in the right insert of figure~\ref{fig:fig22}(a), where the flow starts to deviate following the curved surface.
It strongly suggests that the rapid shear-layer deviation is crucial for the base pressure recovery.
Considering the transition of $\langle{C_p}\rangle$ in figure~\ref{fig:fig22}(b) when control is turned-off, we note a very rapid decrease of pressure followed by a slow recovery during approximately 5 convective time units ($0.1\,\text{s}$) up to the unforced level, suggesting a period of reestablishment of the initial wake conditions by the detachment of the Coanda surface flow.

\begin{figure}
				        \centering
                \includegraphics[scale=0.38]{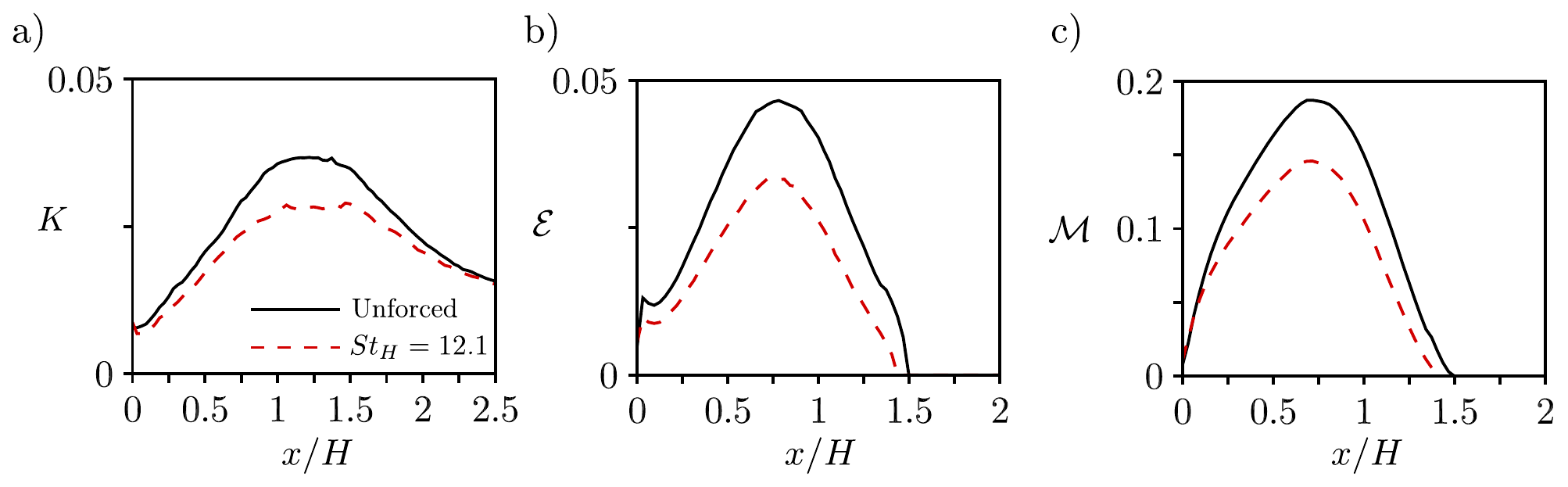}
                \caption{Global measurements in the near wake. Integrated turbulent kinetic energy $K$ (a), total kinetic energy inside the reverse flow region $\mathcal{E}$ (b) and integrated reverse streamwise velocity $\mathcal{M}$ (c). The actuation parameters are $St_H=12.1$ and $C_\mu=7.5\times{10^{-3}}$ with the Coanda effect.}
                \label{fig:fig23}
        \end{figure}

As shown for high-frequency actuation, the smaller shear-layer growth stabilizes the wake velocity fluctuations.
In figure~\ref{fig:fig23}(a), the integrated turbulent kinetic energy ($K$) shows an overall decrease of velocity fluctuations along the wake with unsteady Coanda actuation.
For example, $K$ decreases by more than $20\,\%$ at $x/H\in[0.8,1.5]$.
The total kinetic energy $\mathcal{E}$ in the reverse flow presents a similar behavior (see figure~\ref{fig:fig23}(b)), with reductions of up to $30\,\%$.
Finally, the reverse mass flow $\mathcal{M}$ inside the wake, obtained by integrating:

\begin{equation}
{\mathcal{M}}=\int_{\Omega_{\{\overline{u}<0\}}}{|\overline{u}|}dy,
\end{equation}
is plotted in figure~\ref{fig:fig23}(c).
Part of the calculated decrease of $\mathcal{E}$ comes from the streamwise reverse flow intensity, which is also reduced for this flow.

In summary, the unsteady blowing in combination to the Coanda effect contributes to the base pressure not only by amplifying the shear-layer deviation but also by preserving the stabilizing effect of the shear-layers from high-frequency forcing, resulting in a less-fluctuating wake with lower entrainment.

\subsection{An evaluation of the flow control energy input}

Of course, a power is needed to generate the pulsed jets capable to reduce drag.
A relation of this power to the recovered energy from drag reduction must be established for flow control purposes.
To this end, a selection of forced wakes is chosen to evaluate the pulsed jet energy and drag variations.
Following the energy analysis discussed in former studies \citep{Freund94,Choi08,Pfeiffer12}, we may define the power ratio $\zeta$ as:

\begin{equation}
\zeta=\frac{\left|{1-\gamma_d}\right|{C_{x_{o}}}S{{U_o}^3}}{{s_j}\overline{{V_j}^3}}.
\end{equation}			
				
It is worth mentioning that the power necessary to maintain the compressed air $P_i$ in the reservoir as well as the electrical energy spent to actuate the valves are not taken into account here.
We are mainly concerned with the jet mechanical energy compared to the recovered drag power.
Table~\ref{tab:table_power} lists the quantities used to compute $\zeta$ for several drag reducing configurations using steady or high-frequency (HF) Coanda blowing when ${Re_{H}}=3.0\times 10^5$.

\begin{table}
\begin{center}
\begin{tabular}{ccccccc}
$St_H \,\text{-Configuration}$ & $V_{j_{\text{eff.}}}\,(\text{m}\text{s}^{\text{-}1})$ &$C_\mu (\times {10^{-3}})$ & $\gamma_p$ & $\gamma_d$ & $\overline{{V_j}^3} (\times{10^3},{\,\text{m}^3\text{s}^{\text{-}3}})$ & $\zeta$   \\
       $0 \,\text{- Steady}$        & 11.5 &7.3   & 0.95 & 0.97  & 1.54 & 1.54   \\
       $0\,\text{- Steady}$         & 15.8 & 13.1 & 0.96 & 0.97  & 4.00 & 0.59   \\
       $12.1\,\text{- HF}$          & 5.6 & 1.7  & 0.85 & 0.9   & 0.23 & 34.5   \\
			 $12.1\,\text{- HF}$          & 11.6 & 7.5  & 0.84 & 0.89  & 2.74 & 3.3    \\
			 $0\,\text{- Steady, Coanda}$ & 11.5 & 7.3  & 0.95 & 0.98   & 1.54 & 1.02   \\
			 $0\,\text{- Steady, Coanda}$ & 15.8 & 13.1 & 0.86 & 0.94  & 4.00 & 1.18   \\
			 $12.1\,\text{- HF, Coanda}$   & 5.6 & 1.7  & 0.80 & 0.89  & 0.23 & 38     \\
			 $12.1\,\text{- HF, Coanda}$   & 11.6 & 7.5  & 0.67 & 0.82  & 2.74 & 5.17   \\
			
  \end{tabular}
  \caption{Power ratio parameter for drag reduction with and without the Coanda effect.}
  \label{tab:table_power}
  \end{center}
\end{table}

Except the high momentum steady blowing, all calculated ratios $\zeta$ are greater than the unity, showing a recovery of energy invested on actuation by drag reduction.
Generally, as expected from the base pressure measurements, the highest efficiency corresponds to the HF forcing configurations in the presence or not of the Coanda surface.
The best compromise between the recovered and actuation power comes with forcing at low momentum coefficient $St_H=12.1$ and $C_\mu=1.7\times{10^{-3}}$: the values of $\zeta$ are 34  and 38 respectively with the use or not of the Coanda effect.
The high values of $\zeta$ are linked to the low exit jet velocities in these configurations, which significantly impact $\zeta$ due to the cubic dependence on $V_j$.
The unsteady Coanda blowing is the configuration with the highest drag reduction of 18\% ($\gamma_d=0.82$), corresponding to an energy saved five times greater than the pulsed jet energy.
Hence, it appears to be a promising strategy for further optimization in view of future applications in bluff body drag reduction. Parametric variations of $r$ at different upstream conditions $U_o$ would bring more insights on the influence of the attachment point along the surface in order to improve drag and power savings.

\section{Concluding remarks}
				
The impact of periodic forcing on the wake and drag of a blunt body is investigated.
Pulsed jets with variable frequency and amplitude are blown at the rear edges of the model and tangentially to the main flow.
Complementary drag and pressure measurements as well as velocity fields acquired from particle image velocimetry reveal mainly two flow phenomena with distinct time scales.

First, an overall decrease of baseline pressure occurs when actuation is applied within a range of frequencies close to the natural wake time scales.
The convection of the pulsed jet structures amplifies the cross-stream dynamics of the shear-layer by the generation of vortical eddies, enhancing the momentum entering the wake and principally shortening the bubble's length.
As a result, the recirculating flow generates lower rear pressure raising the bluff body drag by about $10\,\%$.  

Increasing jet frequency shifts the mixing upstream towards the edges of the model.
In this region, high-frequency forcing induces a deviation of the separated shear-layer, creating what we refer to a \textit{fluidic boat-tailing} effect decoupled from the absolute wake instabilities \citep{Glezer05}. 
A conceptual scenario is proposed to clarify how the free-stream conditions and jet parameters impact this flow deviation.
Higher vectoring angles are measured for lower free-stream velocities, in agreement to the jet vectoring effect reported in \citet{Smith02}.
Moreover, it is shown that shear-layer deviation increases with the jet amplitude up to an optimum jet stroke length, independently of the Reynolds number.
Our model suggests a connection between the circulation carried by the vortex core and this optimality, in the light of vortex formation scaling and pinch-off \citep{Dabiri09,Gharib98}, and is similarly supported by recent measurements in a forced axisymmetric wake \citep{Oxlade15}. 

It is additionally found that high-frequency actuation reduces shear-layer growth and dampens velocity fluctuations in the whole wake, resulting in a lower entrainment of external, high momentum fluid in the recirculating bubble.
At a Reynolds number ${Re_{H}}=3.0\times 10^5$, the drag is reduced by $10\,\%$ as a result of the coupled flow deviation and the overall wake stabilization, resulting in a wake length similar to the reference flow.
This is explained by the fact that the reduced velocity fluctuations acts to elongate the bubble length by reducing wake entrainment, while flow deviation tends to decrease the wake width.
The final mean bubble has a higher aspect ratio due to wake shaping, in contrast to the low actuation frequency case which enhances flow fluctuations thus decreasing the recirculating length.
Both scenarios are consistent to the wake equilibrium models proposed by \citet{Roshko55,Roshko93a} some decades ago.

Finally, the addition of the Coanda effect to the actuator system not only preserves the unsteady features of control but also reinforces the flow deviation close to the model.
The resulted adjustment of the pressure gradients along the wake is favorable to the base pressure recovery and decreases the drag by almost $20\,\%$ when actuation at high-frequencies is applied.
The unsteady Coanda blowing analyzed here complements the well-known steady actuation used in road vehicle's drag reduction and paves the way for its future development.
 
In general, the physical mechanisms highlighted here provide some guidelines to forthcoming drag control strategies: systematic variation of the Reynolds number, the jet slit thickness, the blowing angle and the Coanda geometry will help on the scaling laws necessary to further applications in real road vehicles \citep{Seifert15}.
To conclude, we believe these results additionally improve our understanding on how bluff body drag varies with wake forcing, which is crucial to find out novel control strategies and implementation, as the recent applications in feedback systems \citep{Brunton15}.

\section*{Acknowledgements}
We are deeply indebted for indispensable experimental support of J.M. Breux, F. Paill\'{e}, R. Bellanger and P. Braud and for stimulating discussions with V. Parezanovi\'{c} and R. Li.
The thesis of D.B is supported financially by PSA - Peugeot Citro\"{e}n and ANRT in the context of the OpenLab Fluidics between the Institut Pprime and PSA - Peugeot Citro\"{e}n (fluidics@poitiers).
We thank J. \"{O}sth and S. Krajnovi\'{c} for supporting our experimental work
with illuminating LES data of the same set-up. 
The authors thank the funding of the Chair of Excellence - Closed-loop control of turbulent shear flows using reduced-order models (TUCOROM)- supported by the French Agence Nationale de la Recherche (ANR).
Last, but not least, we acknowledge the many insightful suggestions from three anonymous referees leading to a substantial improvement of the manuscript.

\section*{Appendix A. Increase of drag by low-frequency actuation}

In order to better understand how low forcing frequency increases drag, we focus on actuation at $St_H=0.8$ with jet amplitude $V_{j_{\text{eff.}}}\sim7.3\,\text{m}\text{s}^{\text{-}1}$.
This configuration leads to a $9\,\%$ raise of drag at $Re_H=3\times{10^5}$, associated to a $12\,\%$ drop of base pressure. 

\begin{figure}
				        \centering
                \includegraphics[scale=0.44]{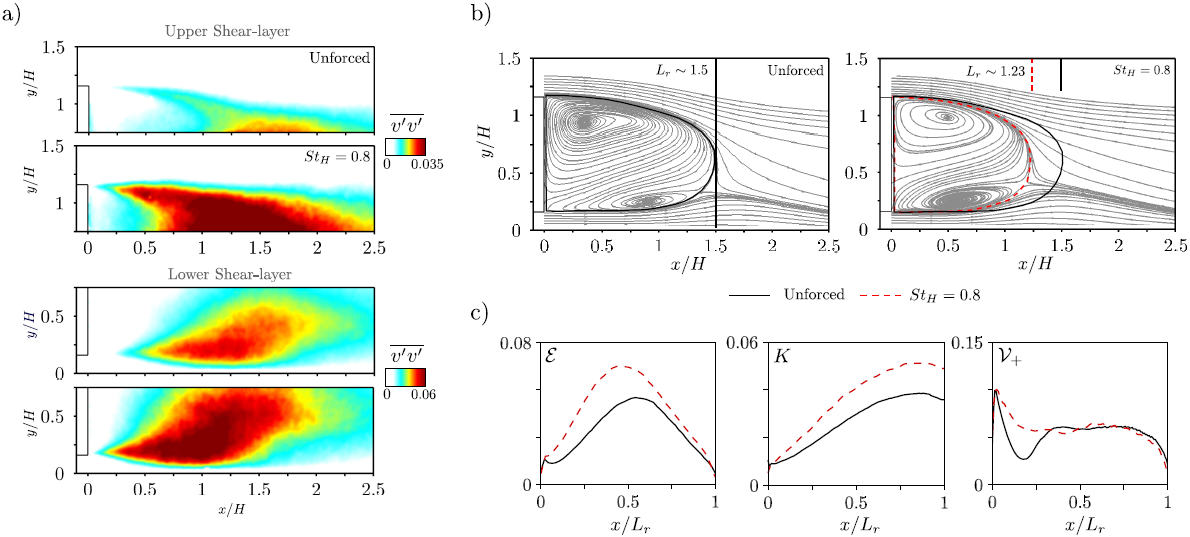}
                \caption{Cross-stream velocity fluctuations, mean flow streamlines and entrainment measures. a) Contour maps of $\overline{v'v'}$ comparing the unforced and actuated flow at a forcing frequency $St_H=0.8$ with $V_{j_{\text{eff.}}}\sim7.3\,\text{m}\text{s}^{\text{-}1}$. b) Velocity streamlines indicating the reduction of the recirculating flow length $L_r$ under actuation. c) Streamwise evolution of ${\mathcal{E}}$, $K$ and ${\mathcal{V}_{+}}$ showing an increase of wake entrainment.}
                \label{fig:fig24}
        \end{figure}

Contrary to the high-frequency actuation, in which flow fluctuations along the entire bubble are damped, we note a significant increase of velocity unsteadiness displayed in figure~\ref{fig:fig24}(a).
The increase of $\overline{v'v'}$ occurs earlier along the lower and upper shear-layers, and extends up to the end of the recirculating flow.

The enhancement of cross-flow dynamics points to a raise of momentum crossing the bubble.
In figure~\ref{fig:fig24}(b), a shortening of the recirculating flow of $18\,\%$ confirms the higher entrainment.
Although our actuation is the same along all the edges, vertical flow asymmetry is still present due to the ground proximity, attested previously by the base pressure distribution in figure~\ref{fig:fig8}(b).

We proceed similarly as for the high-frequency case by computing the integral quantities ${\mathcal{E}}$, $K$ and ${\mathcal{V}_{+}}$.
Their streamwise evolution is plotted in figure~\ref{fig:fig24}(c). 
The mean kinetic energy in the reverse flow significantly raises and its maximum value is located close the center of the bubble ($x/{L_r}\sim0.5$) and increased by more than $30\,\%$ compared to the natural flow.
Part of this comes from an amplification of the turbulent kinetic energy along the whole wake, which continually grows up to $x/{L_r}\sim1$ and is significantly higher for the forced configuration.
Finally, the integrated cross-stream velocity in the reverse flow indicates a net increase of recirculating intensity from the values of ${\mathcal{V}_{+}}$ close to the rear surface up to $x/{L_r}\sim0.3$.      

\begin{figure}
				        \centering
                \includegraphics[scale=0.42]{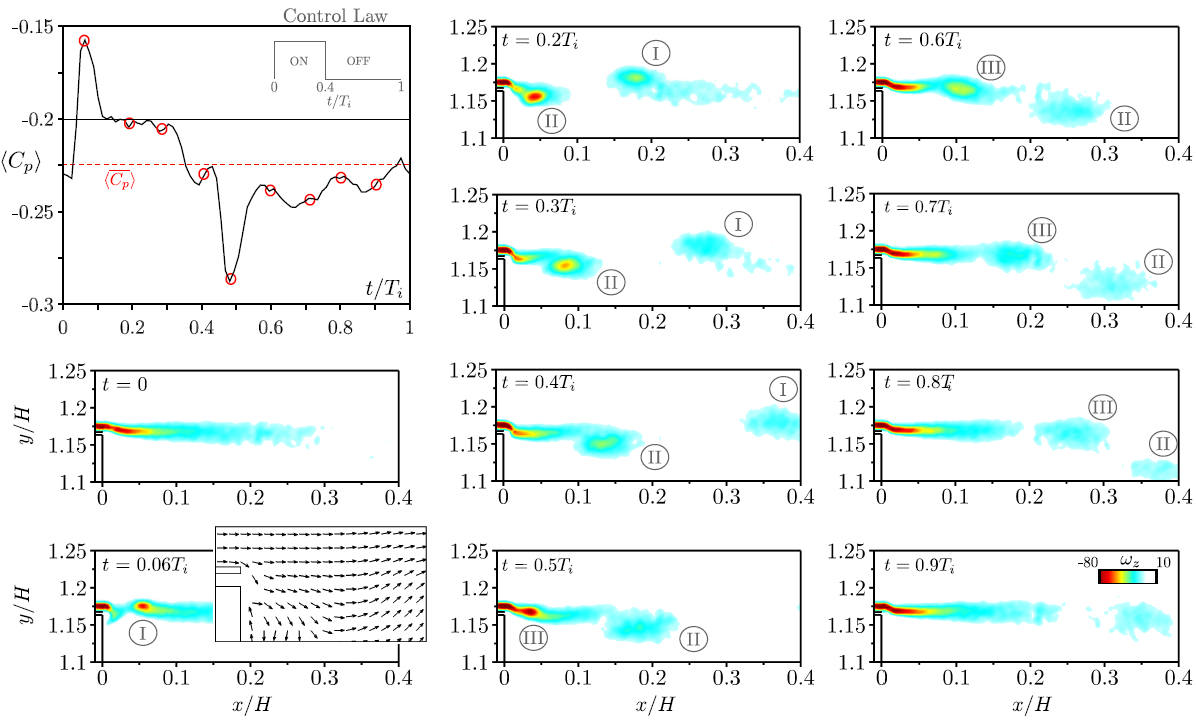}
                \caption{Phase-locked base pressure $\langle{C_p}\rangle$ and vorticity $\omega_z$ field of the upper shear flow when actuation is set at the forcing frequency $St_H=0.8$ and $V_{j_{\text{eff.}}}\sim7.3\,\text{m}\text{s}^{\text{-}1}$. The averaged pressure coefficient for the reference and actuated flows are respectively indicated by the full and dashed lines. The vorticity snapshots correspond to the phases indicated in the pressure plot, respectively $t=\{0, 0.06, 0.2, 0.3, 0.4, 0.48, 0.6, 0.7, 0.8, 0.90\}{T_i}$. The presence of a detached vortex from the initial disruption of the vorticity sheet (I), the main pulsed vortex head (II) as well as the trailing roll-up (III) can be identified by the vorticity contours maps. Velocity vectors indicating the change of the velocity field curvature during the formation of (II) are depicted at $t=0.06{T_i}$. }
                \label{fig:fig25}
        \end{figure}

The pulsed jet evolution in the vicinity of the edges enlightens the interaction mechanisms with the shear-layer and its mixing enhancement.
To correlate the impact of the jet structures with the base pressure, figure~\ref{fig:fig25} presents averages for in-cycle $t/{T_i}$ phases containing the vorticity ($\omega_z$) contours of the pulsed jet and the corresponding base pressure values.
The valve opening starts at $t/T_i=0$, as detailed in the insert by the rectangular control waveform.
During the initial stroke phase, a sudden increase of $\langle{C_p}\rangle$ occurs, followed by a gradual decrease of base pressure   until the valve closing ($t/T_i=0.4$).
The base pressure significantly decreases up to ($t/T_i\sim0.5$), and then successively recovers its mean value $\langle\overline{C_{p}}\rangle$, represented by the dashed line.

A first phase-locked snapshot illustrates the vorticity sheet of the separated boundary layer prior to the jet emission $t/T_i=0$.
The very beginning of the pulsed jet formation induces a disruption of this vortex layer ($t=0.06{T_i}$), leading to the roll-up of a clockwise rotating structure (I), which is afterwards convected downstream.
At this phase, a noteworthy increase of base pressure is measured, more than $20\,\%$ greater than $\langle\overline{C_{p}}\rangle$.
The origin of this sudden pressure raise might be related to the instantaneous change of flow curvature close to the rear edges at the beginning of the cycle (the mean flow, however, does not show a fluidic boat-tailing effect as in the high-frequency case).
The velocity vectors in $t=0.06{T_i}$ show a wave-like flow topology due to the roll-up of the main vortex core, whose formation and convection can be better seen at $t=0.2{T_i}$ and $t=0.3{T_i}$ (II). 

A further decrease of pressure is distinguished close to the end of the stroke cycle at $t=0.40{T_i}$ and immediately after ($t=0.50{T_i}$), when a trailing clockwise vortex (III) is formed due to the sudden deceleration created by closing the solenoid valve.
Both the trailing vortex and the main vortex head are convected and the pressure gradually raises up to $\overline{C_{p_T}}$ during $t=\{0.60,\,0.7,\,0.8,\,0.9\}{T_i}$.
The present high-speed PIV field of view does not allow to study the interactions of structures (II) and (III) further downstream.
However, measurements at higher actuation frequencies ($St_H=1.4$ and $St_H=2.0$) reveals amalgamation of theses structures in this measurement domain (see also \citet{Chaligne13b}).

Since unsteady forcing amplifies the velocity fluctuations, it is worth to associate this dynamics to the time scales of the shear-layer instabilities.
In $\S4.2$, it was shown that $St_\theta$ based on the separating boundary-layer does not scale the broadband range of frequencies around $St_H\sim0.8$.
On the other hand, one may speculate if the jet structures excite the shear-layer (SL) modes further downstream.
To test this hypothesis, we extract the mean streamwise velocity at $x/H\sim0.4$ and approximate it to a hyperbolic tangent profile with equivalent maximum peak vorticity.
Following \citet{Ho84}, we estimate the most amplified frequency for turbulent profiles using $St_\theta\sim0.022-0.024$ and obtain $f_{\text{SL}}\sim St_H\sim0.85$, in good agreement with $St_H\sim0.8$.
By considering its local vorticity thickness $\delta_\omega$, we obtain $St_{\delta_{\omega}}\sim0.09$ corresponding well to 0.098 found by \citep{Morris03}.  

This estimate nicely represents the actuation time scale when compared to the local most amplified shear-layer frequencies at $x/H\sim0.4$.
Our analysis is supported by past findings from \citet{Oster82}, in which is shown that excitation of shear-layers with frequencies smaller than the initial most amplified time scales impacts the dynamics further downstream.
However, given the actuation magnitude and its well formed coherent structures, one may wonder if the natural instabilities still dominate the shear development under forcing.
If it is not the case, the above estimates would be misleading and nonlinear effects implied by actuation should be further considered.

\bibliographystyle{jfm}
\bibliography{References}

\end{document}